\documentclass[traditabstract]{aa}

\usepackage{graphicx}
\usepackage{natbib}
\bibpunct{(}{)}{;}{a}{}{,} 

\usepackage{amsmath,amssymb}
\usepackage{url}
\usepackage{fixltx2e}

\usepackage{txfonts}

\newcommand{\msun}{\ensuremath{\mathrm{M}_{\sun}}}

\begin{document}

\title{Accuracy and efficiency of raytracing photoionisation algorithms}
\author{Jonathan Mackey\thanks{Alexander von Humboldt Fellow}}
\institute{Argelander-Institut f\"ur Astronomie, Auf dem H\"ugel 71, 53121 Bonn, Germany.\quad \email{jmackey@astro.uni-bonn.de}}

\date{Received August 2011 / Accepted XXX}

\abstract{
Three non-equilibrium photoionisation algorithms for hydrodynamical grid-based simulation codes are compared in terms of accuracy, timestepping criteria, and parallel scaling.  Explicit methods with first-order time accuracy for photon conservation must use very restrictive timestep criteria to accurately track R-type ionisation fronts.  A second-order accurate algorithm is described which, although it requires more work per step, allows much longer timesteps and is consequently more efficient.  Implicit methods allow ionisation fronts to cross many grid cells per timestep while maintaining photon conservation accuracy.  It is shown, however, that errors are much larger for multi-frequency radiation than for monochromatic radiation with the implicit algorithm used here, and large errors accrue when an ionisation front crosses many optical depths in a single step.  The accuracy and convergence rates of the different algorithms are tested with a large number of timestepping criteria to identify the best criterion for each algorithm.  With these criteria selected, the second-order explicit algorithm is the most efficient of the three, and its parallel scaling is significantly better than that of the implicit algorithm.  The upgrade from first- to second-order accuracy in explicit algorithms could be made very simply to fixed-grid and adaptive mesh-refinement codes which currently use a first-order method.
}

\maketitle

\section{Introduction}
\label{sec:intro}
Photoionisation is an important process in many situations in astrophysics, being the dominant energy source driving the dynamics of H~\textsc{ii} regions around massive stars \citep[e.g.][]{MatODel69}.
Numerical simulations of photoionisation proceeded from the pioneering 1D models of \citet{Las66a} to early 2D calculations in the 1980s \citep*[e.g.][]{BodTenYor79,SanWhiKle82}; numerical methods and results from these early works were reviewed in detail by \citet{Yor86}.
Various degrees of approximation have been employed, for example assuming ionisation equilibrium \citep*[e.g.][]{GarSegFra96}, non-equilibrium monochromatic radiation \citep*[e.g.][]{WhaAbeNor04}, or non-equilibrium multi-frequency radiation \citep[e.g.][]{FraMel94}.
3D calculations on Cartesian grids became possible in the late 1990s \citep{RagMelArtEA99,AbeNorMad99} and later on grids using 3D adaptive mesh refinement (AMR) \citep{LimMel03}, 2D static nested grids \citep*{FreHenYor03} and for 3D smoothed particle hydrodynamics (SPH) \citep[e.g.][]{KesBur03}.

Computing limitations have dictated that most calculations where the microphysics is coupled to the dynamics have used the on-the-spot (OTS) approximation for diffuse radiation.
This assumes that recombinations to the ground state do not change the ionising radiation field because the photon produced by this recombination takes the place of a similar photon that will be absorbed rapidly to reionise the recombined atom.
The OTS approximation therefore ignores both the change in energy and in direction of the recombination radiation; in addition, it becomes significantly more complicated when helium is also included in the model.
Its validity and effects are a subject of ongoing research \citep[e.g.][]{Rit05,RagHenVasEA09,WilHen09}.
More complicated schemes that do include recombination radiation are now being developed for multi-dimensional codes, such as the non-ionising radiative transfer of \citet{KuiKlaDulEA10} and \citet{ComHenAudEA10}, the cosmological reionisation \textsc{radamesh} scheme of \citet{CanPor11}, and the Monte-Carlo photoionisation algorithm of \citet{HawHar12}.

This work focusses on comparing methods using the OTS approximation, which generally fall into two major categories: explicit schemes requiring short timesteps and implicit schemes with less restrictive timesteps.
These are introduced and discussed separately in the following paragraphs.
The terms ``explicit'' and ``implicit'' here refer to the raytracing algorithm and hence to the inputs to the microphysics integration, not to the actual method used to integrate microphysical quantities within a cell (which is usually at least partially implicit).
An explicit scheme therefore uses an instantaneous optical depth for the ray entering a cell, whereas in an implicit scheme the optical depth contains information from both the initial and time-advanced solution.

\subsection{Explicit timestepping schemes}
The photoionisation method implemented by \citet{WhaNor06} in a version of the \textsc{zeus-mp} code is representative of many algorithms commonly used in astrophysics fluid dynamics codes: first the timestep is calculated, followed by a hydrodynamics update and then an operator-split source term update (the order of the hydrodynamics and source term updates can be interchanged).
\citet{WhaNor06} also include substepping of the chemistry reaction network within the source term evaluation because the chemical timescales can be much shorter than the dynamical timescales.
For models with radiative transfer, each substep involves calculating the optical depth to every grid point followed by a time integration of the rate equations and internal energy equation, and as such is first-order accurate in time for photon conservation.
The substep timestep criterion used was rather stringent: $\Delta t=0.1 n_e/\dot{n}_e$, where $n_e$ is the electron number density and $\dot{n}_e$ its partial time derivative.
The method of \citet{KruStoGar07} for photoionisation in the \textsc{athena} code uses a similar time-integration scheme.
\citet{MacTorOisEA07} implemented the \citet{AbeNorMad99} method in the \textsc{zeus-mp} code to study H~\textsc{ii} region expansion with a simple heating and cooling implementation. \citet{WisAbe11} describe a photoionisation module for the AMR code \textsc{enzo}, using similar chemistry and timestepping routines to \citet{WhaNor06} but with some modifications to make the scheme more efficient.
As described in the literature, all of these methods are first-order accurate in time as regards photon conservation, and also use very restrictive timestepping criteria.

\citet{RijPleDubEA06} presented a photoionisation algorithm written for the \textsc{flash} code (Hybrid Characteristics -- \textsc{flash-hc}) that demonstrated good parallel scaling and accuracy in calculating optical depths.
It did not restrict the timestep by the photoionisation time, thereby significantly reducing the computational cost of a calculation, but at the expense of propagating R-type ionisation fronts too slowly \citep{IliCiaAlvEA06}.
To alleviate this problem in the code tests of \citet{IliCiaAlvEA06}, an extra timestep restriction was imposed that was triggered by the presence of R-type ionisation fronts, although the form of this restriction was not specified.
Building on this work, \citet{PetBanKleEA10} significantly improved the \textsc{flash-hc} algorithm and used it to model massive star formation and the growth of H~\textsc{ii} regions around young stars while they are still accreting gas from their surroundings.
Although it is not described in \citet{PetBanKleEA10}, the \textsc{flash-hc} integration should be second-order accurate in time because the source terms are evaluated both in the half step and the full step of the hydrodynamics update (T.\ Peters, private communication).
This gives a time-centred density field for the raytracing in the full step update, significantly increasing the accuracy of the method, as will be demonstrated here.
This scheme appears to avoid the timestepping limitations of other explicit algorithms by sacrificing some accuracy in the tracking of R-type ionisation fronts.

Explicit schemes scale reasonably well on parallel clusters because so little computation is required in the raytracing step, but the sequential nature of the raytracing is always a limiting factor.
For problems with a simple spherical geometry, reasonable parallel scaling can also be obtained by only decomposing the domain along surfaces of constant angle \citep{WhaNor06}.
The main limitation of explicit algorithms, however, is that the timesteps must be so short -- for optically thick neutral cells an ionisation front can only cross at most a single cell per microphysics integration, leading to a Courant-like limit applied to the ionisation front velocity.

\subsection{Implicit timestepping schemes}
The C$^2$-ray method \citep{MelIliAlvEA06} is an implicit raytracing algorithm that was designed to overcome the timestepping restrictions of fully explicit schemes.
During the tracing of rays outwards from the source, cells are integrated forwards a full timestep and a time-averaged optical depth (or attenuation fraction in the case of \citealt{MacLim10}) is passed to the next cell.
The optical depths to every cell therefore contain information from both the initial and final states and are hence implicit.
This allows ionisation fronts to cross many optically thick grid cells per timestep, an impossibility for an explicit scheme.
This method has been shown to conserve photons well and to track R-type ionisation fronts with the correct speed, as long as the timestep is limited to a fraction of the recombination time \citep{MelIliAlvEA06,IliCiaAlvEA06,MacLim10}.
It was designed primarily for calculation of the reionisation of the universe by the first stars and protogalaxies \citep[e.g.][]{IliMelPenEA06}, but has been successfully coupled to hydrodynamics (HD) and magnetohydrodynamics (MHD) codes to model both the expansion of H~\textsc{ii} regions into turbulent density fields around single massive stars \citep{MelArtHenEA06,ArtHenMelEA11}, and the photoionisation of dense globules \citep{HenArtDeCEA09,MacLim10,MacLim11b}.
The only real weakness of this method as a time integration scheme is that the microphysics integration happens during the raytracing step, which must be performed in sequence outwards from the sources and hence has rather poor scaling properties on distributed memory computing clusters.
It is therefore more suited to shared memory systems.

The main aim of this work is to compare the accuracy and parallel scaling of explicit and implicit schemes implemented with the same code and to identify a sufficient timestep criterion for explicit schemes to accurately track R-type ionisation fronts.
The code and algorithms used here are described in Sect.~\ref{sec:alg}.
The accuracy of the three algorithms is compared in Sect.~\ref{sec:acc} for the case of monochromatic ionising radiation, and in Sect.~\ref{sec:multifrequency} for multi-frequency ionising radiation with frequency-dependent optical depths.
The parallel scaling of the algorithms is assessed in Sect.~\ref{sec:scaling} for static and dynamical situations.
The results are discussed and conclusions are summarised in Sect.~\ref{sec:conc}.

\section{Algorithm implementation}
\label{sec:alg}
The algorithms tested here have been implemented in the raytracing/photoionisation/magnetohydrodynamics (R-MHD) code described in \citet{MacLim10,MacLim11b}, to which the reader is referred for further details.
It is a finite-volume, uniform-grid code written in \texttt{C++} and parallelised by domain decomposition, with communication of internal boundary data through the message passing interface (MPI).
The equations of inviscid compressible HD or ideal compressible MHD are solved on a uniform fixed grid in 1-3D; here a spherically symmetric 1D grid, a plane-parallel 1D grid, an axisymmetric 2D grid in $(z,R)$, and a 3D Cartesian grid are used.
These equations are supplemented by a microphysics integrator for the ion fraction of Hydrogen, $y$, and a raytracer to calculate column densities from either point sources or from sources at infinity.
A tracer variable is used for $y$ (and any other species to be integrated); this tracer is passively advected with the flow and also has creation (ionisation) and destruction (recombination) source terms.
Microphysical (radiative and collisional) heating and cooling processes also provide a source term to the energy equation.
For HD the equations solved are as follows (the conservation of mass, momentum, energy, and H$^+$ ions, respectively):
\begin{align}
  \frac{\partial\rho}{\partial t} + \nabla\cdot[\rho\mathbf{v}] &= 0 \nonumber \\
  \frac{\partial\rho\mathbf{v}}{\partial t} +\nabla\cdot[\mathbf{v}\otimes\rho\mathbf{v}]
+\nabla p_{\mathrm{g}}&=0 \nonumber\\
  \frac{\partial E}{\partial t} +\nabla\cdot\left\{\mathbf{v}\left[E+p_{\mathrm{g}}\right]\right\} &=
    \Gamma(\rho,y,N_{\mathrm{H}},N_{\mathrm{H0}}) - \Lambda(\rho,y,T) \nonumber\\
  \frac{1}{\rho}\left\{\frac{\partial [\rho y]}{\partial t} +\nabla\cdot[\rho y\mathbf{v}]\right\} &= A_{\mathrm{pi}}(\rho,y,N_{\mathrm{H0}})[1-y]+\nonumber\\
    A_{\mathrm{ci}}(T)&n_{\mathrm{H}}y[1-y] -\alpha_{\mathrm{rr}}^{\mathrm{B}}(T)n_{\mathrm{H}}y^2  \,.
\label{eqn:Euler}
\end{align}
Here the gas density, pressure, velocity, and total energy density are $[\rho,p_{\mathrm{g}},\mathbf{v},E]$ respectively, where $E=E_{\mathrm{int}}+0.5\rho v^2$ is the sum of internal and kinetic energy densities.
For a gas with constant adiabatic index $\gamma$, we have $E_{\mathrm{int}}=p_{\mathrm{g}}/[\gamma-1]$.
The total H number density is $n_{\mathrm{H}} = \rho /(2.4\times10^{-24} \,\mathrm{g})$.
$\Gamma$ and $\Lambda$ are the heating and cooling rates per unit volume in the cell ($\mathrm{erg}\,\mathrm{cm}^{-3}\,\mathrm{s}^{-1}$), respectively.
$\Lambda$ is a function of $\rho$, $y$, and the gas temperature $T$, 
while $\Gamma$ depends also on the column density of H nucleons, $N_{\mathrm{H}}$, and of neutral H, $N_{\mathrm{H0}}$.
The collisional ionisation ($A_{\mathrm{ci}}$) and Case B radiative recombination ($\alpha_{\mathrm{rr}}^{\mathrm{B}}$) rates are functions of $T$ and are in units of cm$^3\,$s$^{-1}$;
the photoionisation rate ($A_{\mathrm{pi}}$) is a function of $(\rho,y,N_{\mathrm{H0}})$ and distance from the source, with units of $\mathrm{s}^{-1}$.

The homogeneous parts of these equations are integrated using a directionally unsplit, second-order (in time and space), finite volume formulation described for axisymmetry by \citet{Fal91} and for Cartesian geometry by \citet{FalKomJoa98}.
The scheme for spherical coordinates in 1D is a trivial modification of the axisymmetric algorithm; some results from \citet{BosMyh92} were used for the second-order reconstruction.
Microphysical source terms are then solved by operator splitting using one of three possible algorithms, described in the following subsections.
Photoionisation and ionisation-heating rates require the optical depth and distance from any radiation sources to the cell in question.
This is calculated using a short characteristics ray-tracing module with the interpolation weighting scheme advocated by \citet{MelIliAlvEA06} (this is more accurate than the weighting proposed in appendix B of \citealt{RijPleDubEA06}); diffuse radiation is treated approximately by the OTS approximation.
In the microphysics update the source terms are integrated, giving the following ODEs:
\begin{align}
\dot{y} &= A_{\mathrm{pi}}(\rho,y,N_{\mathrm{H0}})[1-y]+A_{\mathrm{ci}}(T) n_{\mathrm{H}}y[1-y]  -\alpha_{\mathrm{rr}}^{\mathrm{B}}(T)n_{\mathrm{H}}y^2  \,\nonumber \\
\dot{E}_{\mathrm{int}} &= \Gamma(\rho,y,N_{\mathrm{H}},N_{\mathrm{H0}}) - \Lambda(\rho,y,T)  \,.
\label{eqn:source_terms}
\end{align}
Here the density is constant for each cell so temperature is a function only of $E_{\mathrm{int}}$ and $y$.
All algorithms described below use the same microphysics integrator and heating and cooling rates to enable a fair comparison between models.
Variables are integrated in time using backward differencing with Newton iteration, implemented with the \textsc{cvode} solver of the \textsc{Sundials} numerical integration library~\citep{CohHin96}.

The heating and cooling functions use either the \citet{MacLim10} model C2 or the much more detailed model of \citet{HenArtDeCEA09}, which was calibrated using a dedicated photochemistry code (although here their X-ray heating term is omitted).
The more detailed model enables the inclusion of multi-frequency photoionisation sources (to model the spectral hardening of radiation with optical depth) and heating due to far-ultraviolet (FUV) non-ionising stellar radiation, both of which have a significant effect on photoionisation simulations.
It also provides a more realistic cooling function for dense neutral gas, although the details of the cooling physics are not so important for this work.
The code can be switched by a compile flag to use either the C2 heating/cooling function with monochromatic radiation, or the more detailed heating/cooling with multi-frequency radiation.
The multi-frequency photoionisation and photo-heating rates are pre-calculated for a given source spectrum and tabulated as a function of optical depth as described in e.g.~\citet{FraMel94} and \citet{MelIliAlvEA06}.

\subsection{Implicit algorithm}
The raytracing/microphysics scheme used in \citet{MacLim10,MacLim11b} is a variant of the C$^2$-ray algorithm \citep{MelIliAlvEA06}, and will be referred to here as Algorithm 1 (or simply A1).
Some improvements have been made to the algorithm, so it is described again here.
The algorithm has two interfaces with the main simulation code: one for calculating the simulation timestep and one for updating the microphysical quantities.
In each timestep, first the timestep $\Delta t$ is calculated, then the combined raytracing and microphysics update of the internal energy density ($E_{\mathrm{int}}$) and neutral fraction ($1-y$) is performed, followed by a second-order-accurate dynamics update.
The timestep criteria are discussed in more detail below, but a basic requirement for accurate tracking of R-type ionisation fronts is that the timestep must be limited to a fraction of the recombination time, $t_{\mathrm{rec}}=1/\alpha_{\mathrm{rr}}^{\mathrm{B}}n_{\mathrm{H}}$,~\citep{MelIliAlvEA06}.

For A1 the two source term integrations defined by Equations~\ref{eqn:source_terms} are supplemented by integrating the attenuation along the ray segment passing through the cell as described in \citet{MacLim10}.
This allows the calculation of a time-averaged attenuation fraction, which can be converted to a time-averaged column density.
The integration is performed at the ionisation threshold $h\nu_0=13.6\,$eV, and the time-averaged attenuation fraction of photons at $\nu=\nu_0$ is then
\begin{equation}
\langle f_{\nu_0}\rangle=
\frac{1}{\Delta t}\int_t^{t+\Delta t} \exp[-\Delta\tau_{\nu_0}(t^\prime)] dt^\prime \,,
\end{equation}
where the cell optical depth, $\Delta\tau_\nu = n_{\mathrm{H0}} \Delta s \sigma_\nu$ is the product of the neutral H number density, the ray segment length $\Delta s$, and the photoionisation cross-section $\sigma_\nu$.
This time-averaged attenuation fraction is converted to a time-averaged column density and used to calculate the column density to the next cell further from the source.

It was found to be more numerically stable to integrate the neutral fraction than the ion fraction, so the three variables integrated are $(1-y, E_{\mathrm{int}}, \exp[-\Delta\tau_{\nu_0}] )$.
The variables are integrated using \textsc{cvode} with a relative error tolerance of $10^{-4}$ and with an absolute error tolerance of $10^{-12}$, $10^{-17}$, and $10^{-30}$, respectively.
This is a more accurate integration scheme than that used in \citet{MacLim10,MacLim11b} and is consequently more computationally expensive.

\begin{figure}
\centering
\resizebox{0.95\hsize}{!}{\includegraphics{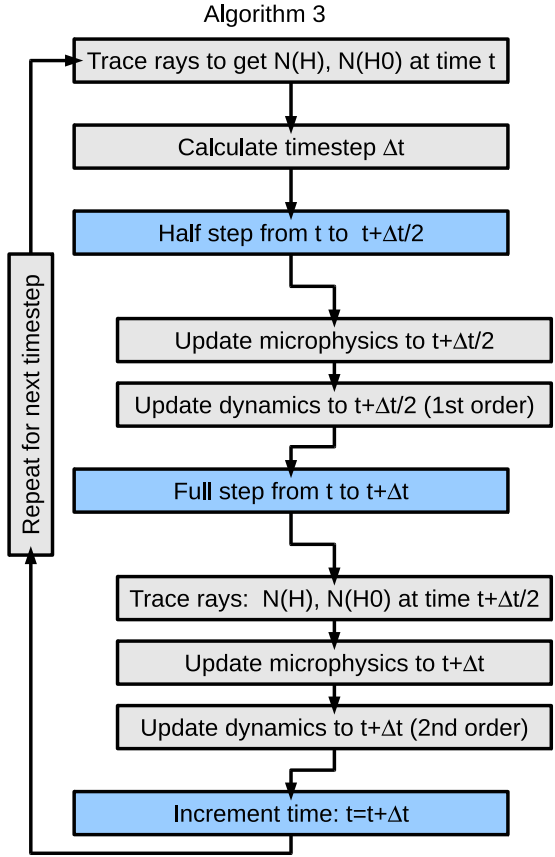}}
\caption{Sequence of calculations for a single timestep for algorithm 3, which traces rays twice per step: the time-centred second raytracing makes photon conservation second-order in time, thereby allowing a less restrictive timestep and consequently proving to be more computationally efficient than algorithm 2.  Second-order accurate microphysics substeps could use the same algorithm, but omitting the dynamics updates.}
\label{fig:Algorithm3}
\end{figure}

\subsection{Explicit algorithms}

Two explicit integration algorithms have been implemented, the first of which is similar to previously published methods (see section~\ref{sec:intro}).
Algorithm 2 is a replacement of A1 with a new timestep criterion and microphysics integration, similar to the \citet{WhaNor06} algorithm, but without substepping.
It is again fully operator-split from the dynamics and a timestep proceeds as follows:
\begin{enumerate}
\item Rays are traced to calculate neutral (and optionally total) column densities to each cell.
\item Microphysical and dynamical timesteps are calculated; the minimum over all cells is used.
\item Microphysical quantities are integrated from $t\rightarrow t+\Delta t$ using the instantaneous column densities.
\item Using this intermediate state as a starting point, a second-order dynamics update is performed over the time interval $\Delta t$.
\end{enumerate}
The key difference between A2 and A1 is that instantaneous column densities are used from the beginning of the timestep for the integration of the microphysics equations.
This allows the raytracing step to be separated from the microphysics update, which has parallel scaling advantages already discussed.
The microphysics integration is exactly the same as for A1 except that the time-averaged attenuation fraction is not needed, so only two variables are integrated, i.e.~$(1-y,E_{\mathrm{int}})$.
It was found that an accurate solution with A2 required very restrictive timesteps, which would make multi-dimensional calculations very computationally expensive (timestepping criteria are discussed in the next section).
This is because the optical depths are not time-centred in a timestep and so photon conservation is first-order accurate in time, making it very difficult to accurately track R-type ionisation fronts.

Algorithm 3 is a modification of A2 in which raytracing is performed twice per step, once at the beginning to calculate the timestep, and secondly using the time-centred half-step density (and ion fraction) field.
The sequence of calculations for A3 is shown in Fig.~\ref{fig:Algorithm3}.
The second raytracing sets the optical depths used for the full step microphysics update and leads to second-order accurate photon conservation.
While it requires significantly more calculation per step compared to A2 (two raytracings and source term integrations per step), the higher order of accuracy leads to a much more computationally efficient integration, as will be shown in the next section.
During the writing of this paper it was discovered (T.\ Peters, private communication) that A3 is similar to the time update used in the upgraded version of \textsc{flash-hc}, although here the timestepping criterion has been varied to track R-type ionisation fronts accurately.

\subsection{Timestepping criteria} \label{ssec:timestepping}
Previous authors using algorithms similar to A2 and A3 have employed a wide variety of different timestepping criteria, so an attempt is made here to identify a sufficient criterion for each algorithm.
For A1 the timestep should be a fraction of the recombination time, so a constant $K_1$ is defined by $\Delta t = K_1t_{\mathrm{rec}}$.
A timestep limited by the relative change in energy is also considered ($\Delta t=K_2E_{\mathrm{int}}/|\dot{E}_{\mathrm{int}}|$), and by the relative or absolute change in $y$ ($\Delta t=K_3\max(0.05,1-y)/|\dot{y}|$ and $\Delta t=K_4/|\dot{y}|$, respectively).
If all four criteria were used together, the microphysics timestep limit would be given by
\begin{equation}
\Delta t=
  \min\left(K_1t_{\mathrm{rec}},\;K_2\frac{E_{\mathrm{int}}}{|\dot{E}_{\mathrm{int}}}|,\;
  K_3\frac{\max(0.05,1-y)}{|\dot{y}|},\;K_4\frac{1}{|\dot{y}|}\right) \,.
\end{equation}
The constant 0.05 in the third criterion is required to prevent the timestep from becoming very small when a cell approaches full ionisation.
Different combinations of these criteria are assigned an ID number and listed in Table~\ref{tab:timestepping}.
For criteria dt00-dt04 the explicit integration is limited only by the absolute change in $y$, and the implicit algorithm only by the recombination time.
For criteria dt05-dt08 all algorithms are limited additionally by the fractional change in energy, and the implicit algorithm by the absolute change in $y$ rather than the recombination time.
For criteria dt09-dt12 the limit on the absolute change in $y$ is replaced by a limit on the relative change in $(1-y)$, i.e.\ the neutral H fraction.

\begin{table}
  \centering
  \begin{tabular}{| l |c c c c|c c c c|}
    \hline
      & \multicolumn{4}{|c|}{Implicit} & \multicolumn{4}{|c|}{Explicit} \\
    \hline\hline
    ID & $K_1$ & $K_2$ & $K_3$ & $K_4$ & $K_1$ & $K_2$ & $K_3$ & $K_4$ \\ 
    \hline\hline
    dt00 & 1   & $\infty$ & $\infty$ & $\infty$ & $\infty$ & $\infty$ & $\infty$ & 1 \\
    dt01 & 0.3   & $\infty$ & $\infty$ & $\infty$ & $\infty$ & $\infty$ & $\infty$ & 1/2 \\
    dt02 & 0.1   & $\infty$ & $\infty$ & $\infty$ & $\infty$ & $\infty$ & $\infty$ & 1/4 \\
    dt03 & 0.03  & $\infty$ & $\infty$ & $\infty$ & $\infty$ & $\infty$ & $\infty$ & 1/8 \\
    dt04 & 0.003 & $\infty$ & $\infty$ & $\infty$ & $\infty$ & $\infty$ & $\infty$ & 1/16 \\
    \hline
    dt05 & $\infty$ & 1/2  & $\infty$ & 1/2  & $\infty$ & 1/2  & $\infty$ & 1/2 \\
    dt06 & $\infty$ & 1/4  & $\infty$ & 1/4  & $\infty$ & 1/4  & $\infty$ & 1/4 \\
    dt07 & $\infty$ & 1/8  & $\infty$ & 1/8  & $\infty$ & 1/8  & $\infty$ & 1/8 \\
    dt08 & $\infty$ & 1/16 & $\infty$ & 1/16 & $\infty$ & 1/16 & $\infty$ & 1/16 \\
    \hline
    dt09  & $\infty$ & 1/2  & 1/2  & $\infty$ & $\infty$ & 1/2  & 1/2  & $\infty$ \\
    dt10 & $\infty$ & 1/4  & 1/4  & $\infty$ & $\infty$ & 1/4  & 1/4  & $\infty$ \\
    dt11 & $\infty$ & 1/8  & 1/8  & $\infty$ & $\infty$ & 1/8  & 1/8  & $\infty$ \\
    dt12 & $\infty$ & 1/16 & 1/16 & $\infty$ & $\infty$ & 1/16 & 1/16 & $\infty$ \\
    \hline
  \end{tabular}
  \caption{Timestepping criteria used for test simulations.
  Constants $K_1-K_4$ are timestep-limiting factors for four criteria as discussed in section~\ref{ssec:timestepping}.
  Entries with $K_{i}=\infty$ indicate that the criterion is not used.}
  \label{tab:timestepping}
\end{table}

\section{Ionisation fronts for monochromatic radiation}
\label{sec:acc}

The accuracy of the predecessor of A1 was extensively tested in \citet{MacLim10} for monochromatic radiation, and the C$^2$-ray algorithm to which it is closely related has been tested and used successfully in many calculations \citep[e.g.][]{IliWhaMelEA09,ArtHenMelEA11}.
Most emphasis in this section has therefore been devoted to testing the explicit algorithms A2 and A3.
It is important to separate raytracing (spatial discretisation) errors from time integration errors, so tests were performed on a 1D grid, initially in plane-parallel geometry and then in spherical geometry.

\subsection{Plane-parallel radiation without recombinations}
\begin{figure}
\centering
\resizebox{0.85\hsize}{!}{\includegraphics{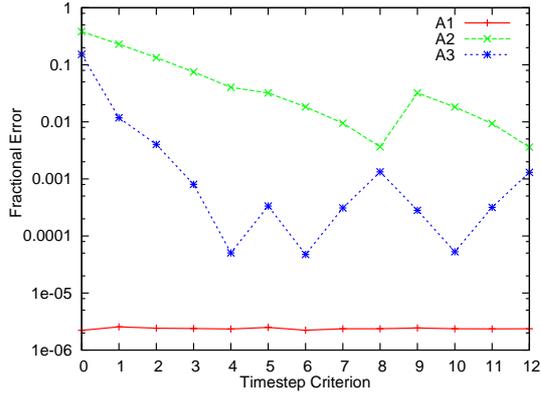}}
\caption{
  Velocity errors as a function of timestep criterion ($x$-axis integers correspond to criteria dt00-dt12) for a 1D plane-parallel ionisation front propagating through a static density field with no recombinations for algorithms A1 (red), A2 (green) and A3 (blue).}
\label{fig:mono_planar}
\end{figure}

The first test is a basic test of photon conservation: a 1D slab-symmetric grid was set up with 100 cells spanning $\simeq0.51\,$pc.
A source was placed at $x=-\infty$ with monochromatic ionising photon flux $F_{\gamma}=10^9\,\mathrm{cm}^{-2}\,\mathrm{s}^{-1}$
  and photon energy $18.6$ eV,
irradiating a neutral uniform density field with $n_{\mathrm{H}}=100\,\mathrm{cm}^{-3}$.
This resulted in an ionisation front propagating with a constant velocity of $v_{\mathrm{IF}}=100\,\mathrm{km}\,\mathrm{s}^{-1}$ through a grid with cell optical depths $\Delta\tau\simeq10$ for a (constant) photoionisation cross-section $\sigma=6.3\times10^{-18} \,\mathrm{cm}^{2}$.
The cross-section was set to be frequency-independent for the monochromatic radiation tests; this was relaxed later for multi-frequency radiation.
For the monochromatic tests in this section the only significance of the cross section is to set the cell optical depth, so it is unimportant that the actual cross section at $18.6$ eV is somewhat smaller than the value used.
Recombinations, collisional ionisation, and dynamics were switched off, so for a perfect integration the number of ions per unit area on the grid is $N_{i}(t)=F_{\gamma}\times t$.
This is a relatively simple problem for A1, but for A2 and A3 the accuracy depends crucially on the timestep criterion because photon attenuation is not averaged over a timestep.

The results are shown in Fig.~\ref{fig:mono_planar} for all three algorithms for all timestep criteria (dt00-dt12).
The relative error is plotted on the $y$-axis as a function of the timestep criterion.
Relative error is defined here as the fractional difference between the number of ions and the number of neutrals, which for this problem corresponds to the fractional error in ionisation front velocity.
There are no recombinations, so the fractional error stays roughly constant for the full simulation, and the plotted values are the mean values for each simulation averaged over many timesteps.
For A1 the error is very small for all timestep criteria, as expected for an algorithm specifically designed to conserve photons.
It is surprising that the error is significantly smaller even than the relative error criterion of the microphysics integrator ($10^{-4}$).

For A2 and A3 the error is very dependent on the timestep criterion, and the much more rapid convergence of A3 with decreasing timestep is clearly shown.
The three different types of criteria can be clearly identified in the A2 curve: dt00-dt04 have the accuracy increasing by roughly a factor of 2 each time, dt05-08 have the same with a smaller initial error, and dt09-dt12 have a similar error.
For A2 the convergence is linear, as expected, but the error is large for all timestep criteria, being significantly less than 1\% only for dt08 and dt12.
With A3, by contrast, dt02 already gives less than 1 per cent error, and convergence is basically quadratic.
For timestep criteria dt05-dt12 the accuracy is within a factor of 10 of the relative error tolerance of the microphysics integrator, so any trends in the error are not significant.

At first glance it seems surprising that A2-dt05 is more accurate than A2-dt04, which has by far the smaller value of $K_4$.
The reason is that dt05 also limits $\Delta t$ by the relative change in energy, so in fact they take almost the same number of timesteps.
The internal energy of a cell increases by a factor of about 320 (from neutral gas with $T\simeq50$ K to ions plus electrons with $T\sim8000$ K) and $K_2=0.5$ allows a 50\% increase in $E_{\mathrm{int}}$ per step, meaning about 14 timesteps are required to go from neutral to ionised.
For dt04 (with $K_4=1/16$) about 16 timesteps are required to ionise a cell, comparable to dt05, and so we expect the two criteria to have a similar accuracy.
Indeed, a log plot of the number of timesteps taken for each criterion with A2 shows approximately the inverse (with arbitrary normalisation) of the fractional error plotted in Fig.~\ref{fig:mono_planar}.

This is a scale-free problem, so the relative error should be independent of ionisation front velocity, and this has been verified numerically for velocities of $v_{\mathrm{IF}}=[10, 30, 100, 300, 1000]\,\mathrm{km}\,\mathrm{s}^{-1}$.
Of course in a dynamical calculation the Courant condition provides an upper limit to the timestep, which will reduce the error for slowly moving ionisation fronts.
For D-type ionisation fronts (subsonic by definition) the Courant condition automatically imposes $K_4<1$.
The simulations were also repeated for densities of $n_{\mathrm{H}}=10\,\mathrm{cm}^{-3}$ and $n_{\mathrm{H}}=1000\,\mathrm{cm}^{-3}$ while keeping the cell optical depth constant, and the results are almost indistinguishable.
The effects of cell optical depth on the accuracy of the three algorithms is studied in more detail for multi-frequency radiation in section~\ref{sec:multifrequency}.

\subsection{Point source radiation in spherical symmetry} \label{sec:mono_spherical}
Here the spherically symmetric expansion of a Str\"omgren sphere in a static medium is calculated and results are compared to the analytic solution.
A 1D spherical grid was set up with 3840 cells uniformly covering the range $r\in[0,1.94]$ pc with a constant gas density of $n_{\mathrm{H}}=100\,\mathrm{cm}^{-3}$ and a temperature $T=50$K, giving a cell optical depth to ionising photons of $\Delta\tau\simeq1$.
A radiation source was placed at the origin with monochromatic ionising photon luminosity $\dot{N}=10^{48}\,\mathrm{s}^{-1}$ and photon energy $18.6$ eV.
The recombination rate of H$^{+}$ was set to be independent of temperature, thereby enabling straightforward comparison with the analytic solution.
Results using densities $10\times$ lower and higher (with associated lower and higher source luminosities) gave almost indistinguishable results.
For simulations with fewer grid cells and correspondingly higher cell optical depth the results are also almost indistinguishable for A1, whereas A2 and A3 become slightly more accurate.
The simulations were again run for all 12 timestep criteria and all three algorithms.

\begin{figure}
\centering
\resizebox{0.85\hsize}{!}{\includegraphics{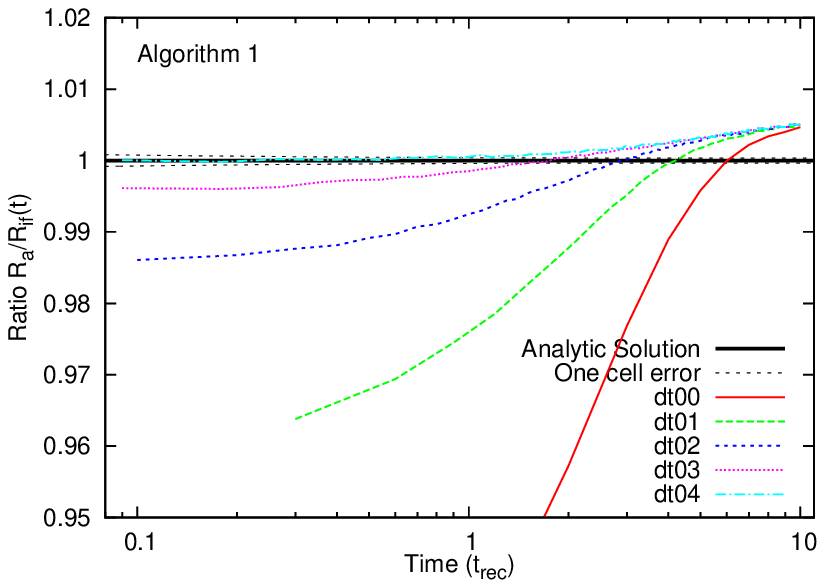}}
\resizebox{0.85\hsize}{!}{\includegraphics{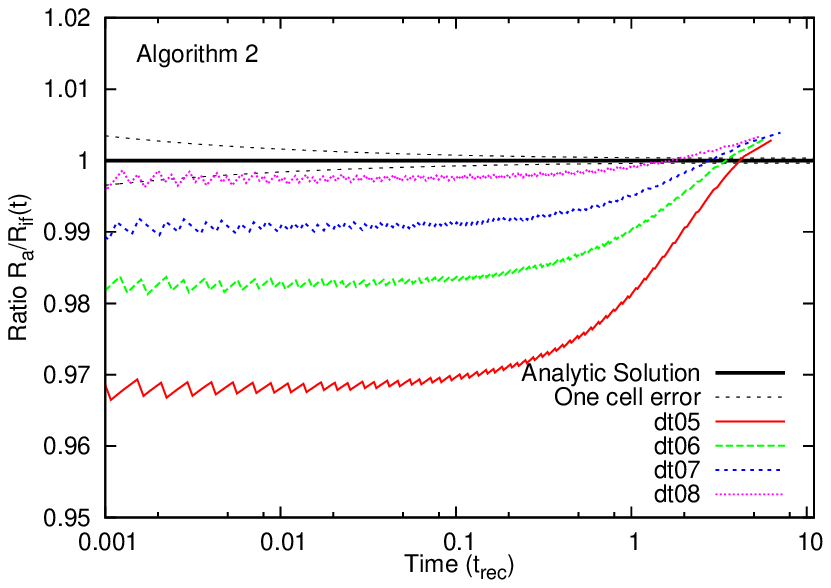}}
\resizebox{0.85\hsize}{!}{\includegraphics{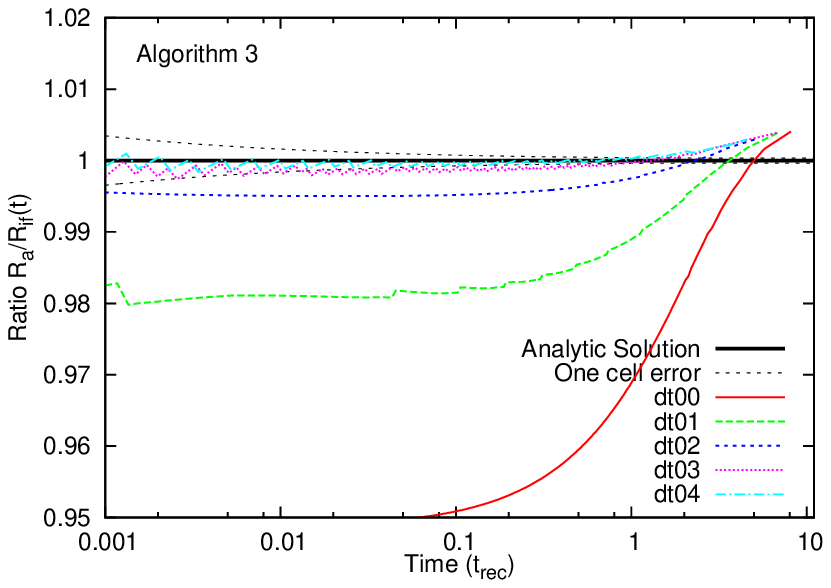}}
\caption{
  Expansion of a Str\"omgren sphere in 1D for the three algorithms: the implicit A1 (top), first-order explicit A2 (centre), and second-order A3 (bottom), with different timestep criteria as indicated.
  The ratio of actual to theoretical ionisation front radius is plotted on a logarithmic time axis from $t=0.001t_{\mathrm{rec}}$ for A2 and A3, and from $0.1t_{\mathrm{rec}}$ for A1 (because the timestep criterion is much less restrictive for A1).
  The different convergence rates for A2 and A3 are again apparent.}
\label{fig:MonoSS1D}
\end{figure}

Sample results for the Str\"omgren sphere expansion are shown in Fig.~\ref{fig:MonoSS1D} for the ratio of the actual ($R_{\mathrm{a}}$) to theoretical ($R_{\mathrm{if}}$) radius given by $R_{\mathrm{if}}(t) = R_S[1-\exp(-t/t_{\mathrm{rec}})]^{1/3}$, where the Str\"omgren radius $R_S$ is here equal to $R_S=1.42$ pc.
Results for dt00-dt04 are shown for A1 and A3, whereas dt05-dt08 are shown for A2 because the results for dt00-dt04 had much larger errors (even dt04 had a 2\% error).
For this calculation, A1 reaches $0.3t_{\mathrm{rec}}$ in a single step for dt01 with an error in position of $\lesssim 4$ per cent, decreasing to $\lesssim1.5$ per cent for $\Delta t=0.1t_{\mathrm{rec}}$ (dt02).  Even at $t=0.1t_{\mathrm{rec}}$ the ionisation front has crossed many cells in a single step with dt02, and the error is smaller than that obtained using A3 with dt01, which takes two timesteps for every grid zone the ionisation front crosses.
This shows the huge efficiency advantages of A1 in tracking R-type ionisation fronts with reasonable accuracy.
If an error of $\lesssim 1-1.5$\% is the maximum allowed, then dt07 is the first timestep criterion that is accurate enough with A2.
With A3 dt02 is already more than accurate enough.
The three simulations with similar accuracy (dt02 for A1, dt07 for A2, and dt02 for A3) took 100, 22661, and 2923 timesteps, respectively, to reach $t=10t_{\mathrm{rec}}$.
In terms of runtime (not counting initialisation and data I/O), the respective runtimes are 12.6, 180, and 65.7 seconds, the differences in runtime being smaller than suggested by the number of timesteps because shorter timesteps have a less costly microphysics integration.

For this problem most of the error accrues at the early stages of expansion ($t\lesssim 0.1 t_{\mathrm{rec}}$) when the ionisation front is R-type and a 10 per cent velocity error can rapidly become a large position error.
The error of course decreases for all timestep criteria when the steady state is approached after a few recombination times.
The steady state does not correspond exactly to the analytic solution because the ion fraction is not a perfect step function in radius -- instead there is a small neutral fraction within the H~\textsc{ii} region that increases as the radius approaches $R_S$~\cite[see e.g.][]{PawSch08}.
This leads to an equilibrium radius slightly larger than $R_S$, with the difference depending on the ionising spectrum.
For the monochromatic radiation used here and the other parameters given above, eq.~33 in \citet{PawSch08} can be solved to show that the equilibrium radius with $y=0.5$ is $r=1.004R_S$, almost exactly the value that the simulations relax to at $t=10t_{\mathrm{rec}}$.
Note that the fractional errors in Fig.~\ref{fig:MonoSS1D} are smaller than for Fig.~\ref{fig:mono_planar} because here the radius corresponds to the cube-root of the number of ions.
There are other physically motivated cases, such as a $1/r^2$ density field, where the ionisation front remains R-type for much longer, and in this situation the ionisation front position errors would continue to increase with time.

\section{Ionisation fronts with multi-frequency radiation} \label{sec:multifrequency}
A more realistic model of propagating ionisation fronts is obtained by considering a spectrum of ionising photons and including the frequency-dependent photoionisation cross-section of H.
The source spectrum is modified as the optical depth increases so there is no guarantee that the results obtained with a monochromatic source will still hold with a multi-frequency source.
To test this, the same calculation as in section~\ref{sec:mono_spherical} was performed using an ionising source spectrum with a blackbody temperature $T=37\,500$ K and normalised to have the same ionising photon luminosity ($\dot{N}=10^{48}\,\mathrm{s}^{-1}$).
The recombination rate was allowed to vary with temperature so the actual Str\"omgren radius is not exactly the same as before.
The number of grid zones was varied to give simulations with cell optical depths (at $\nu=\nu_0$) of $\Delta\tau_0\simeq [1,3,10,30]$ to study the effects of resolution on the solution obtained.
Additionally, the number density of the ISM was varied (with an accompanying change in source luminosity to give the same ionisation front expansion velocity) with $n_{\mathrm{H}}=[10, 100, 1000] \,\mathrm{cm}^{-3}]$.
While the number density had almost no effect on the solution accuracy, it did affect the code efficiency for A1 with dt00-dt04 because the recombination time scales inversely with density.

This gives a grid of 12 simulations, each to be run with 3 different algorithms, each using 12 different timestep criteria.
There are two main considerations for each calculation: the accuracy compared to the most accurate solution, and the runtime.
The best combination of algorithm and timestep criterion will be that which achieves a certain required accuracy with the least computation, with the overall restriction that the computation requirement is not prohibitive.

\begin{figure*}
\centering
\resizebox{0.42\hsize}{!}{\includegraphics{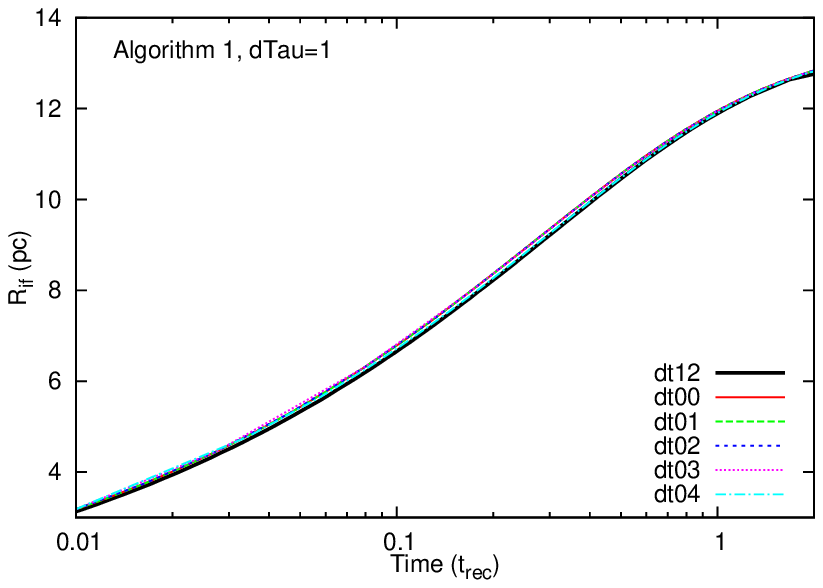}}
\resizebox{0.42\hsize}{!}{\includegraphics{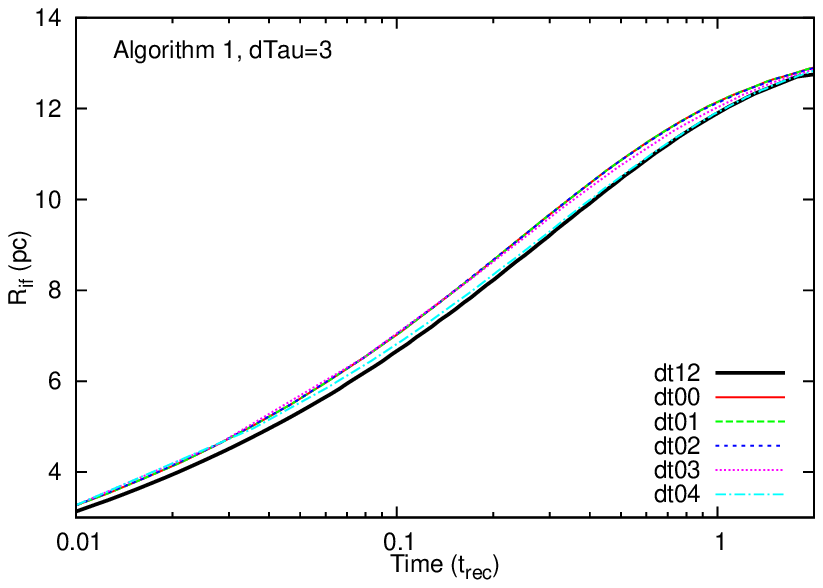}}
\resizebox{0.42\hsize}{!}{\includegraphics{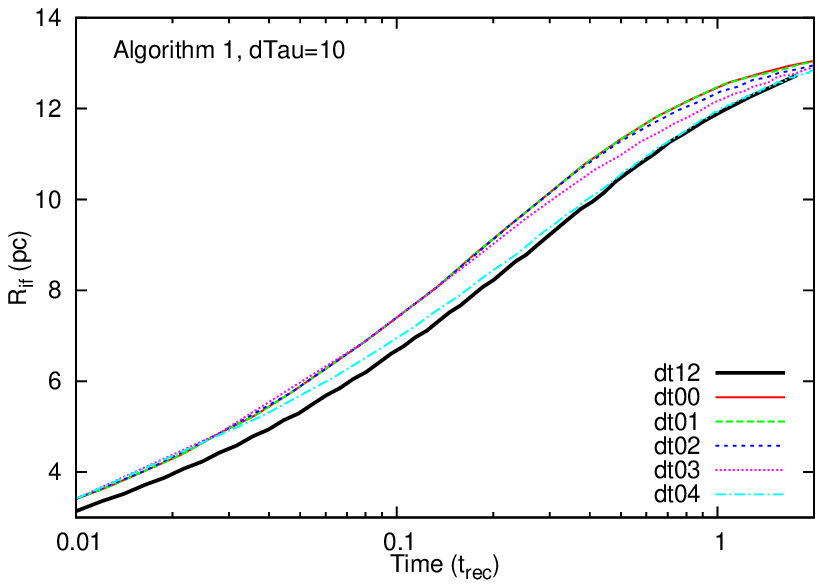}}
\resizebox{0.42\hsize}{!}{\includegraphics{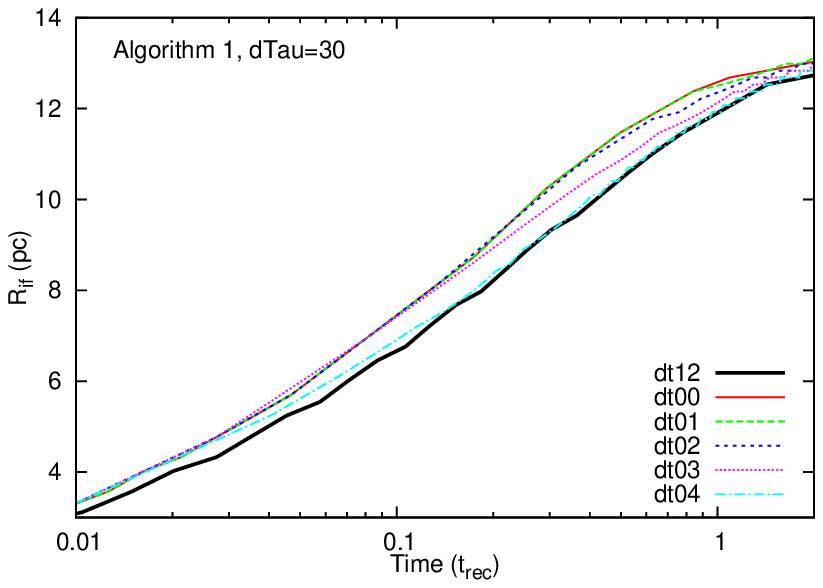}}
\resizebox{0.42\hsize}{!}{\includegraphics{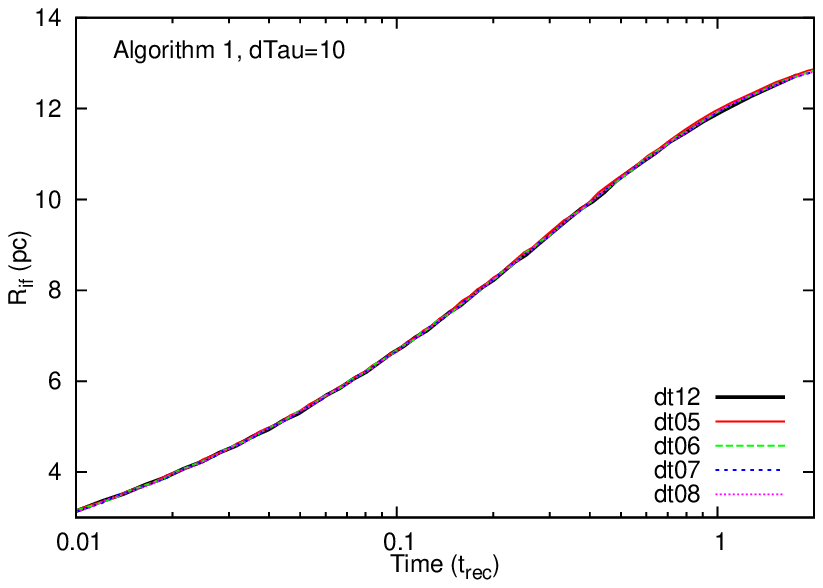}}
\resizebox{0.42\hsize}{!}{\includegraphics{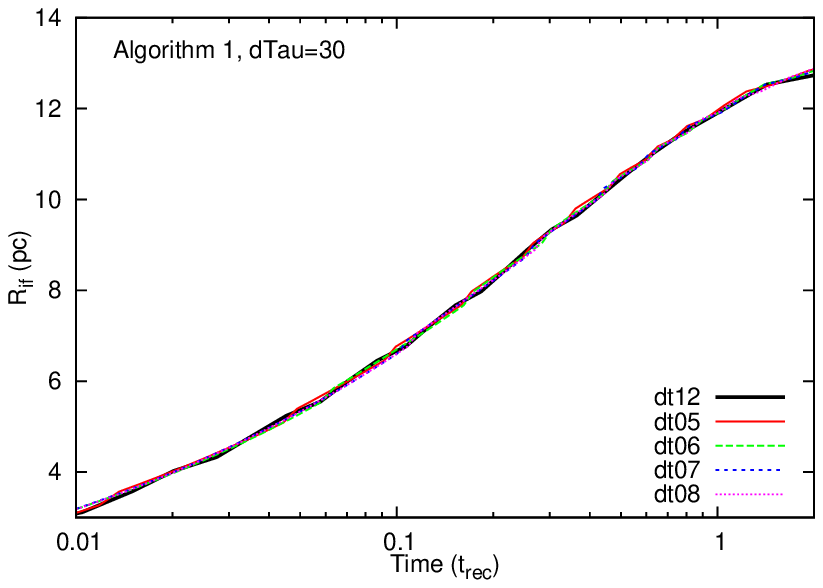}}
\caption{
  Expansion of a Str\"omgren sphere in 1D using A1 with an ambient gas
  density of $n_{\mathrm{H}}=10\,\mathrm{cm}^{-3}$ (results for higher
  densities are almost indistinguishable).  The position of the 
  cell with the steepest radial gradient in H$^+$ fraction is plotted as a
  function of time for simulations with cell optical depth $\Delta\tau_0 \simeq1$ and dt00-dt04 (top left),
  $\Delta\tau_0 \simeq3$ and dt00-dt04 (top right), $\Delta\tau_0 \simeq10$ and dt00-dt04 (centre left),
  $\Delta\tau_0 \simeq30$ and dt00-dt04 (centre right), $\Delta\tau_0 \simeq10$ and dt05-dt08 (below left), 
  and $\Delta\tau_0 \simeq30$ and dt05-dt08 (below right).  The equivalent plots for dt09-dt12 all show
  results indistinguishable from the dt12 curve.  The trend for decreasing accuracy with increasing cell
  optical depth (i.e.\ decreasing numerical resolution) for criteria dt00-dt04 is clearly seen in the
  first four panels; this is corrected by the criteria dt05-dt08 that also limit the timestep by the 
  relative change in internal energy.
  Discreteness in the computational grid and the output frequency account for the non-smooth curves in the plots for $\Delta\tau_0 \simeq10$ and $30$ (also in the following figures).
   }
\label{fig:MFSS1D_radius_A1}
\end{figure*}

\begin{figure*}
\centering
\resizebox{0.42\hsize}{!}{\includegraphics{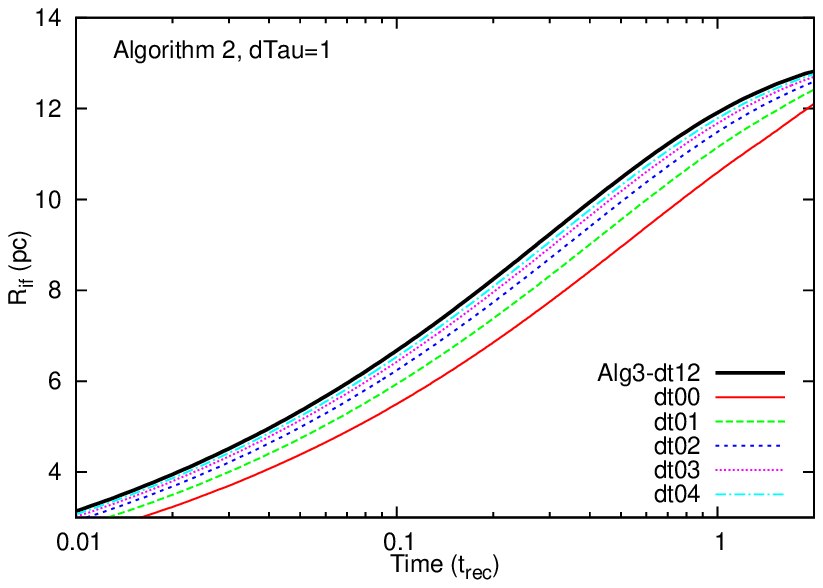}}
\resizebox{0.42\hsize}{!}{\includegraphics{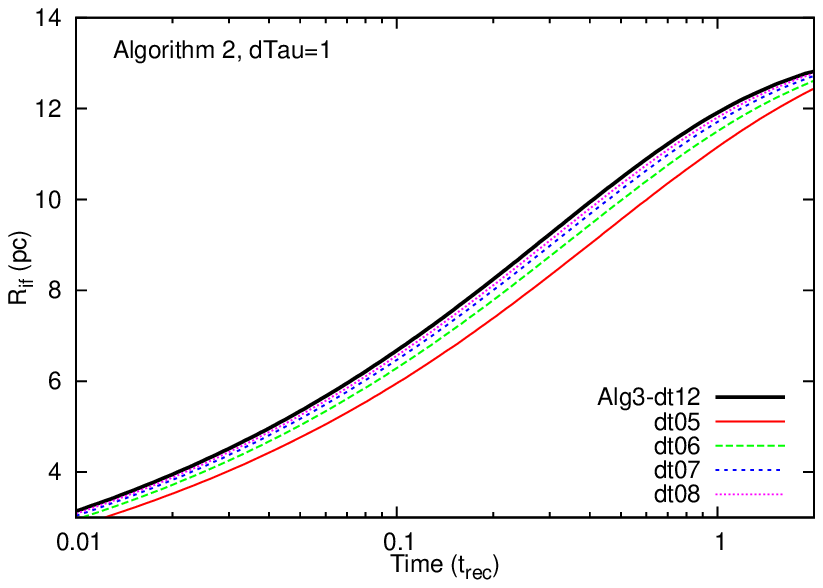}}
\resizebox{0.42\hsize}{!}{\includegraphics{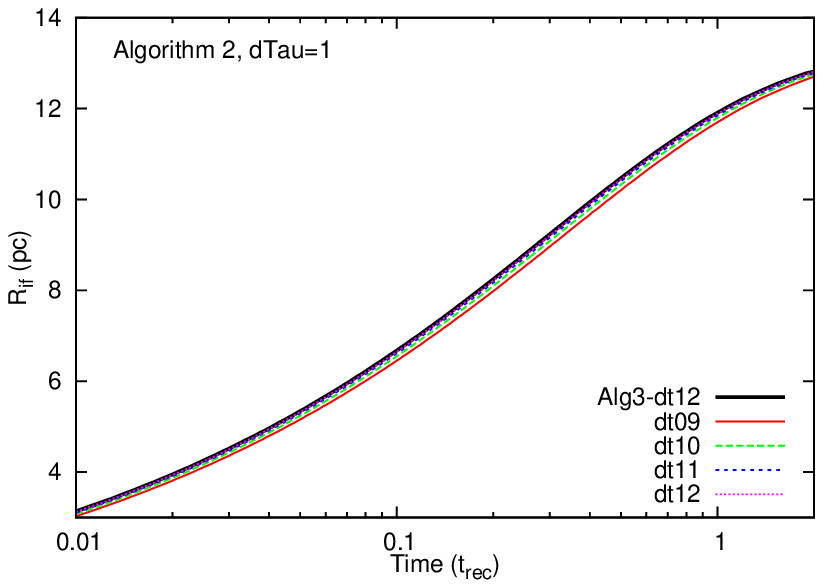}}
\resizebox{0.42\hsize}{!}{\includegraphics{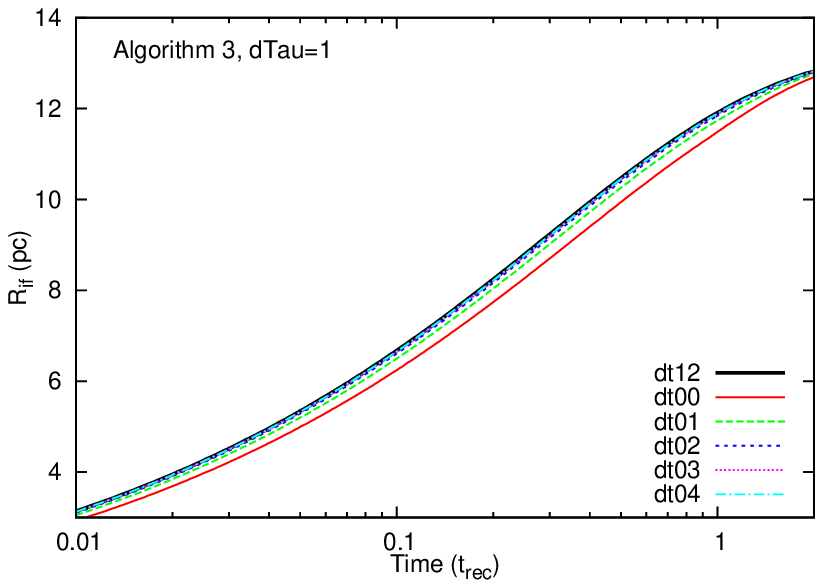}}
\resizebox{0.42\hsize}{!}{\includegraphics{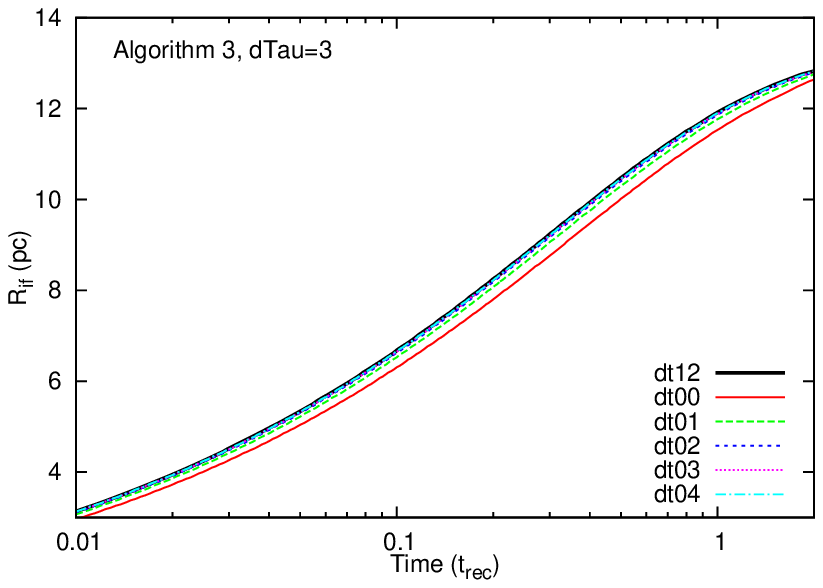}}
\resizebox{0.42\hsize}{!}{\includegraphics{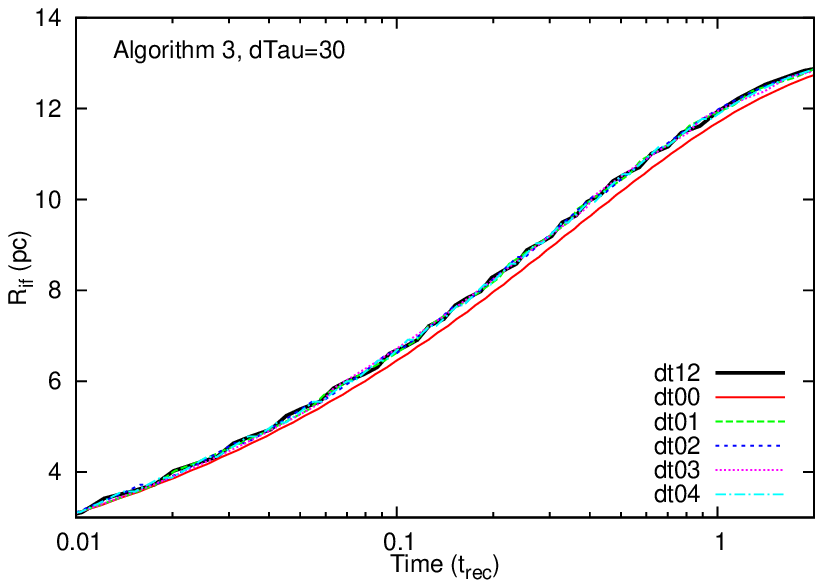}}
\caption{
  As Fig.~\ref{fig:MFSS1D_radius_A1} but for A2 and A3.  The panels show results for 
  $n_{\mathrm{H}}=10\,\mathrm{cm}^{-3}$; results for other ambient densities are indistinguishable, and the accuracy of all criteria increases somewhat with increasing cell optical depth (i.e.\ decreasing numerical resolution).
  The panels show results for A2 with cell optical depth $\Delta\tau_0 \simeq1$ and timestep criteria dt00-04 (top left), dt05-dt08 (top right), and dt09-dt12 (centre left),
  and for A3 with dt00-dt04 and $\Delta\tau_0 \simeq1$ (centre right), $\Delta\tau_0 \simeq3$ (bottom left), and $\Delta\tau_0 \simeq30$ (bottom right).
  In all plots the reference result in the heavy black line is from A3 with dt12.
  }
\label{fig:MFSS1D_radius_A2A3}
\end{figure*}

\begin{figure*}
\centering
\resizebox{0.42\hsize}{!}{\includegraphics{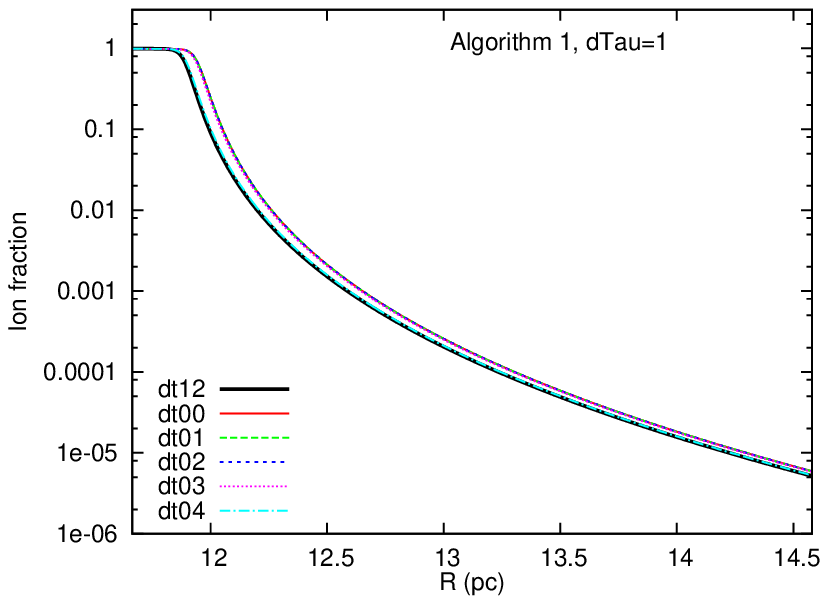}}
\resizebox{0.42\hsize}{!}{\includegraphics{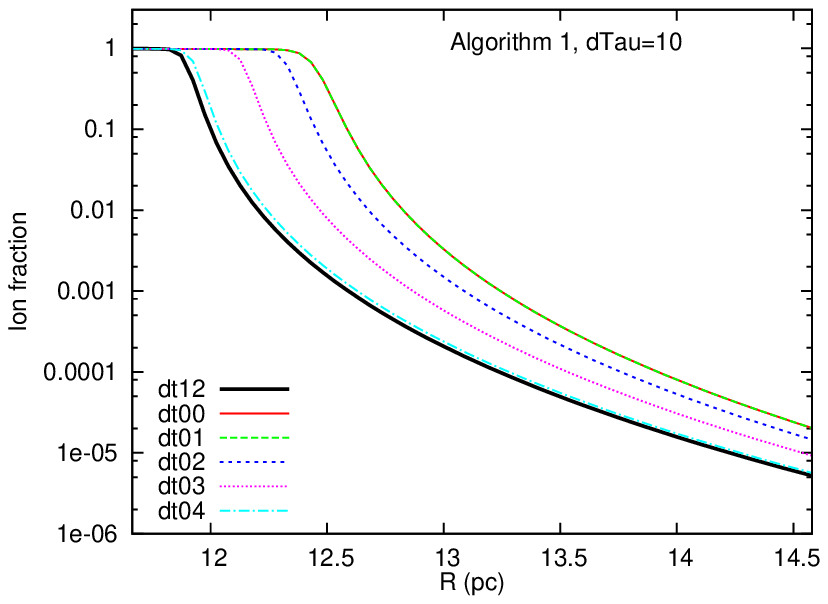}}
\resizebox{0.42\hsize}{!}{\includegraphics{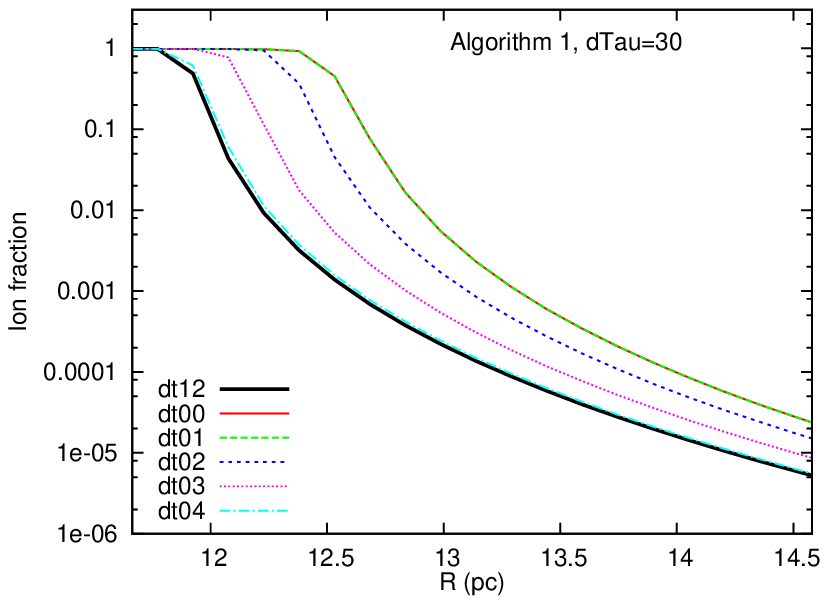}}
\resizebox{0.42\hsize}{!}{\includegraphics{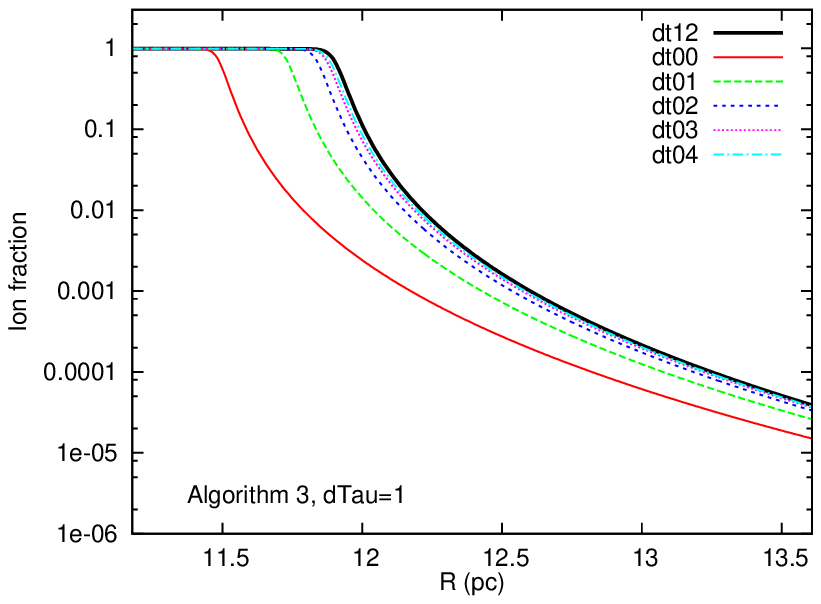}}
\resizebox{0.42\hsize}{!}{\includegraphics{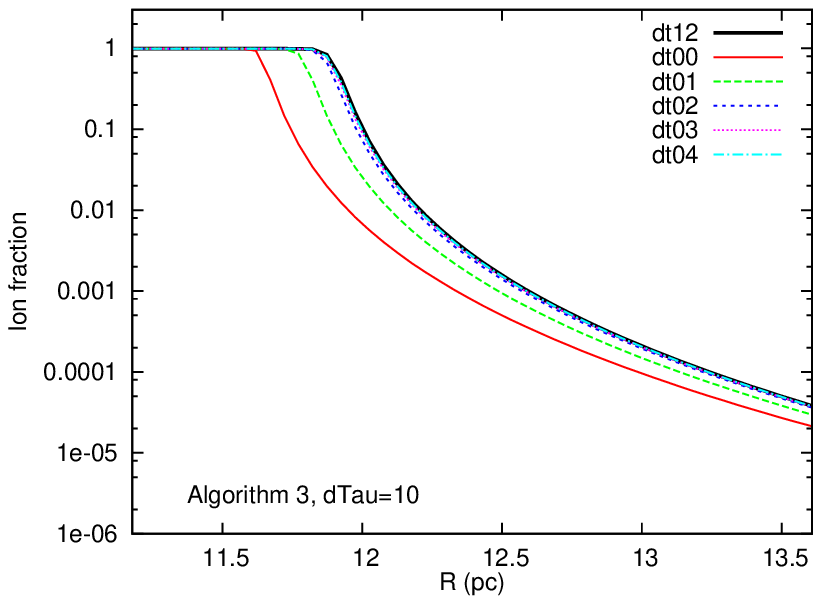}}
\resizebox{0.42\hsize}{!}{\includegraphics{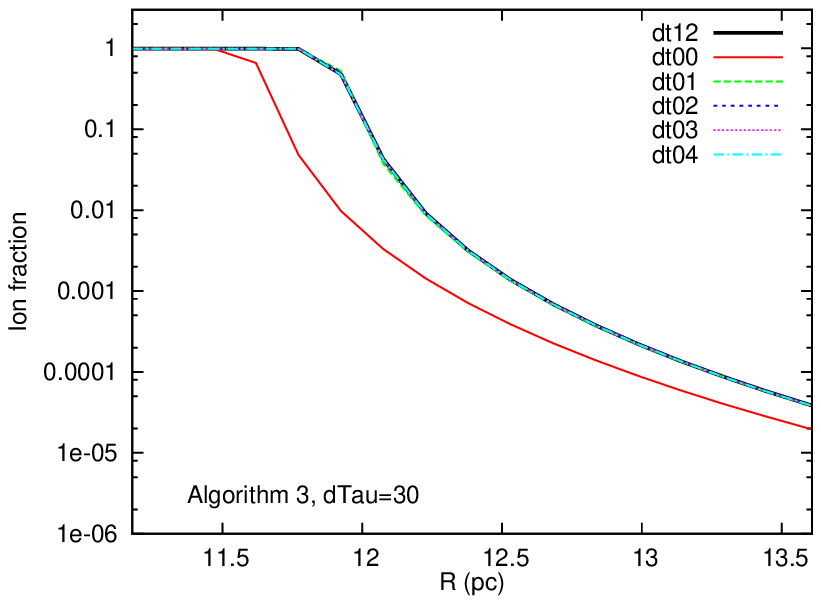}}
\caption{
  Plots of the radial profile of the H$^+$ fraction at $t=t_{\mathrm{rec}}$ for
  A1 (first three plots) and A3 (last three plots).
  Results for $n_{\mathrm{H}}=10\,\mathrm{cm}^{-3}$
  are shown (results for higher densities were indistinguishable) for A1 with cell optical
  depths $\Delta\tau_0 \simeq1$ (top left), $\Delta\tau_0 \simeq10$ (top right),
  and $\Delta\tau_0 \simeq30$ (centre left),
  and the equivalent plots for A3 are centre right, bottom left, and bottom right.
  For A1 and A3 the criteria dt06-dt11 all provide good fits, very close to the dt12 results.
    As in previous figures, the discreteness of the grid is seen in the $\Delta\tau_0 \simeq30$ plots.
  }
\label{fig:MFSS1D_profiles}
\end{figure*}

\subsection{Algorithm accuracy}
The most basic property of the simulation is the location of the ionisation front as a function of time.
This was calculated by finding the cell with the largest second-order radial gradient in the H$^+$ fraction defined by
\begin{equation}
\max\frac{\partial y}{\partial r} = \max\left( \frac{y(r_{i+1})-y(r_{i-1})}{r_{i+1}-r_{i-1}} \right) \quad \forall\; i\in(1,N_i-2)\,,
\end{equation}
where there are $N_i$ grid zones and $i$ is zero-offset.
This produces almost identical results to other criteria, e.g.\ the first cell with $y<0.5$, although as the ionisation front reaches the Str\"omgren radius at $t>t_{\mathrm{rec}}$ the cell with the steepest gradient can occasionally retreat/advance by one cell as the ionisation front relaxes to equilibrium.
Simulations with higher cell optical depths necessarily have fewer and larger cells; grid discreteness effects are clearly seen in Figs.~\ref{fig:MFSS1D_radius_A1} -- \ref{fig:MFSS1D_profiles} for models with cell $\Delta\tau_0 \simeq30$.
The ionisation front radius as a function of time is shown for representative simulations run with A1 in Fig.~\ref{fig:MFSS1D_radius_A1}, and with A2 and A3 in Fig.~\ref{fig:MFSS1D_radius_A2A3}.
Most panels show results for dt00-dt04 compared to the A3 result for dt12 (the most accurate run) because dt00-dt04 use the least restrictive timestep criteria and hence have the largest errors.
Fig.~\ref{fig:MFSS1D_radius_A1} shows that for A1 the best solution is obtained for low cell optical depths, and the error increases steadily with optical depth, to an error in ionisation front radius of about 10-15\% in the worst case.
Data exist for dt00-dt03 at $t\sim0.01t_{\mathrm{rec}}$ with A1 because $\Delta t$ is also limited by the Courant condition, and in addition, the first timestep is artificially set to be very short.
The results show, in contrast to the monochromatic radiation results, that timestep-limiting based only on the recombination time (dt00-dt04) is not reliable for very optically thick cells with multi-frequency radiation.
Timestep-limiting using the relative energy change (dt05-dt08) is much more stringent for the early expansion of the H~\textsc{ii} region and hence provides an accurate solution.
Limiting additionally by the relative change in neutral fraction (dt09-dt12) provides little extra benefit.

In contrast to A1, algorithms A2 and A3 are less accurate at lower cell optical depths.
The first three plots in Fig.~\ref{fig:MFSS1D_radius_A2A3} show results for A2 equivalent to those for A1 in Fig.~\ref{fig:MFSS1D_radius_A1}, except that here only the worst case is shown with cell $\Delta\tau_0 \simeq1$.
Curves for all timestep limiters are also shown, and it can be seen that dt03, dt04, dt07, dt08, and dt09-dt12 provide adequate fits, but only dt10-dt12 are properly converged to the solution obtained with A3.
For A3 only results for dt00-dt04 are shown in the last three plots of Fig.~\ref{fig:MFSS1D_radius_A2A3} because all of dt06-dt12 are indistinguishable from each other on this plot for all densities and cell $\Delta\tau_0$ values.
Only dt00 is a noticeably bad solution with A3.

Fig.~\ref{fig:MFSS1D_profiles} shows the radial profile of the H$^+$ fraction at $t=t_{\mathrm{rec}}$ for A1 (above) and A3 (below) for timestep criteria dt00-dt04, compared to the numerically converged result from dt12.
A1 over-predicts the ionisation front location for low time-accuracy (as seen already in Fig.~\ref{fig:MFSS1D_radius_A1}), indicating that the attenuation of photons is less than it should be.
A3, by contrast, over-attenuates photons for low time-accuracy because it uses instantaneous column densities.
The results for A2 are similar to A3, but the errors are much larger, and the solution converges more slowly because it is a first-order method.
For A1 the dt00 and dt01 results are almost identical, and dt02 also for the $\Delta\tau_0 \simeq1$ simulation, because the Courant timestep condition is more restrictive than the recombination time (the CFL number was set to 1.0 for these tests).
The accuracy is therefore somewhat better than would be obtained without the Courant condition.
When the cell optical depth is low it is clear that A1 is the most accurate algorithm, but the accuracy decreases severely for very optically thick cells using only the recombination times as a limiter (dt00-dt04).
Models dt06-dt12 all produce results similar to A3 with dt06-dt12.
A3 actually gets more accurate for higher optical depths using dt00-dt04, and it can be seen that dt02 provides a good solution in all cases.
Errors in gas temperature are in all cases identical to errors in ion fraction, since both quantities are integrated to the same relative error tolerance.

The reason for the errors in A1 can be traced back to its origin as an algorithm for monochromatic light.
In that case the spectrum cannot change with optical depth and there is a one-to-one correspondence between column density and the fractional attenuation of ionising photons.
Once a multi-frequency source is used, however, the fractional attenuation is a function of the incident spectrum and of the column density, both of which can change with time, meaning that the one-to-one correspondence is no longer so clear.
When the cell optical depth changes significantly during a timestep, the radiation spectrum must also change significantly, so a time-averaged column density at a specific frequency will no longer give an accurate value for the time-averaged photon attenuation fraction.
It is likely that a more accurate algorithm could be devised, and indeed it is possible that the C$^2$-ray algorithm of \citet{MelIliAlvEA06} is already more accurate for this problem than A1.

\subsection{Algorithm efficiency}
\begin{figure*}
\centering
\resizebox{0.42\hsize}{!}{\includegraphics{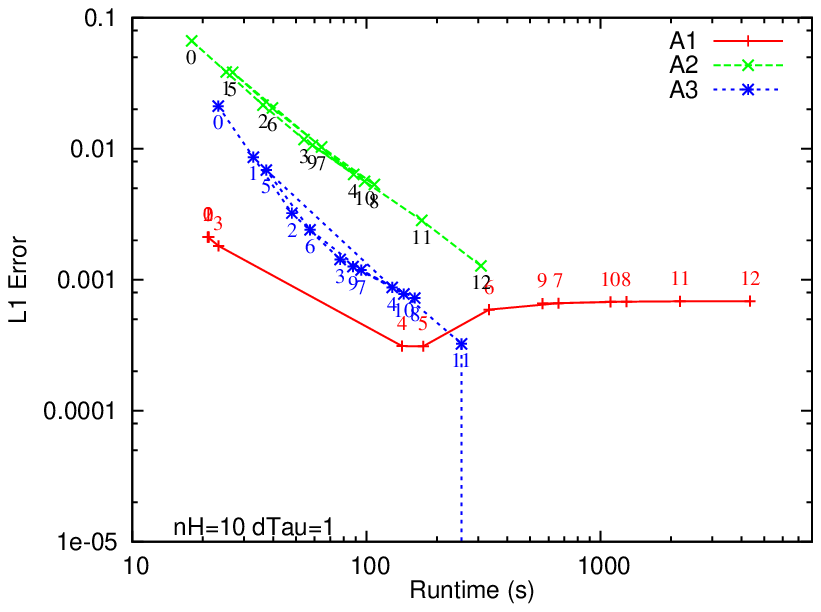}}
\resizebox{0.42\hsize}{!}{\includegraphics{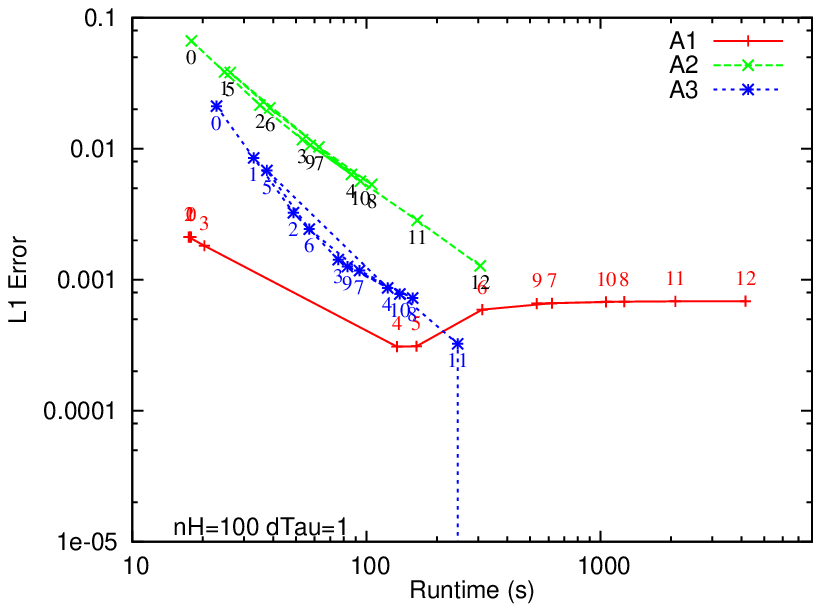}}
\resizebox{0.42\hsize}{!}{\includegraphics{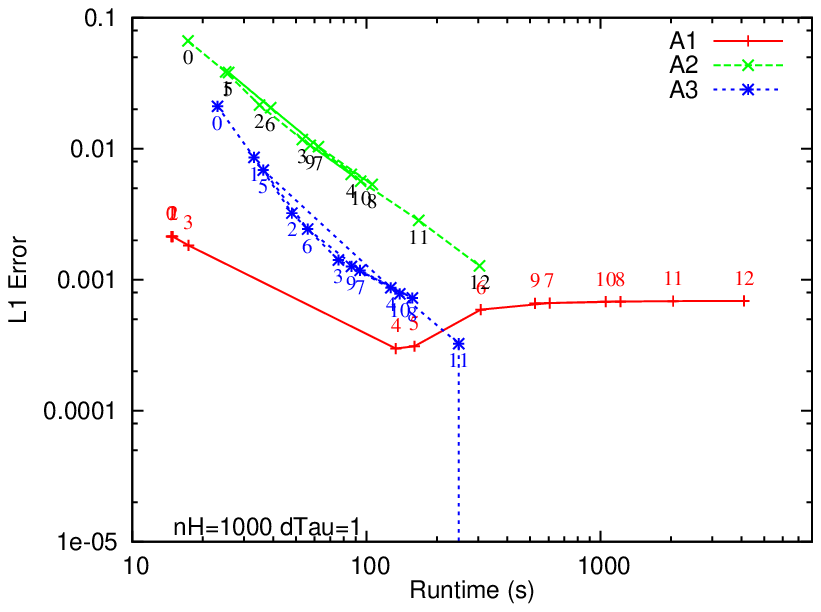}}
\resizebox{0.42\hsize}{!}{\includegraphics{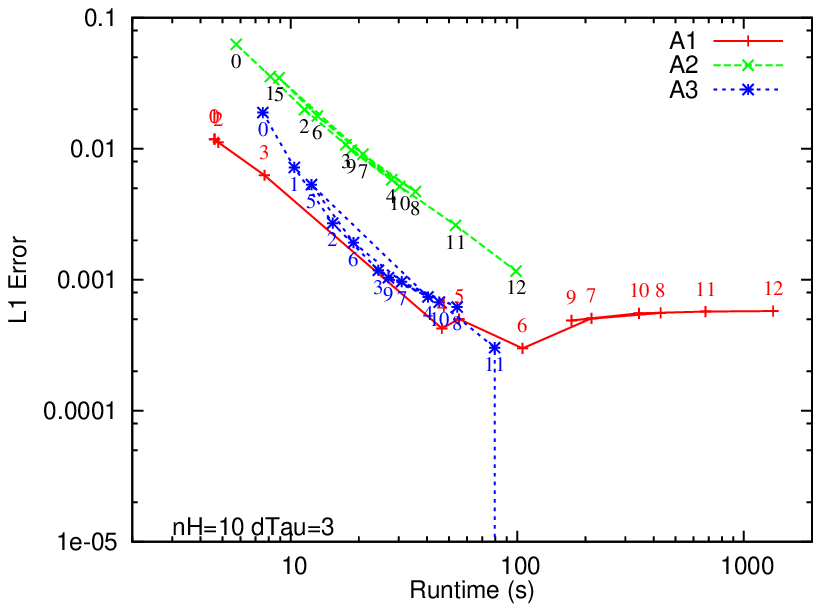}}
\resizebox{0.42\hsize}{!}{\includegraphics{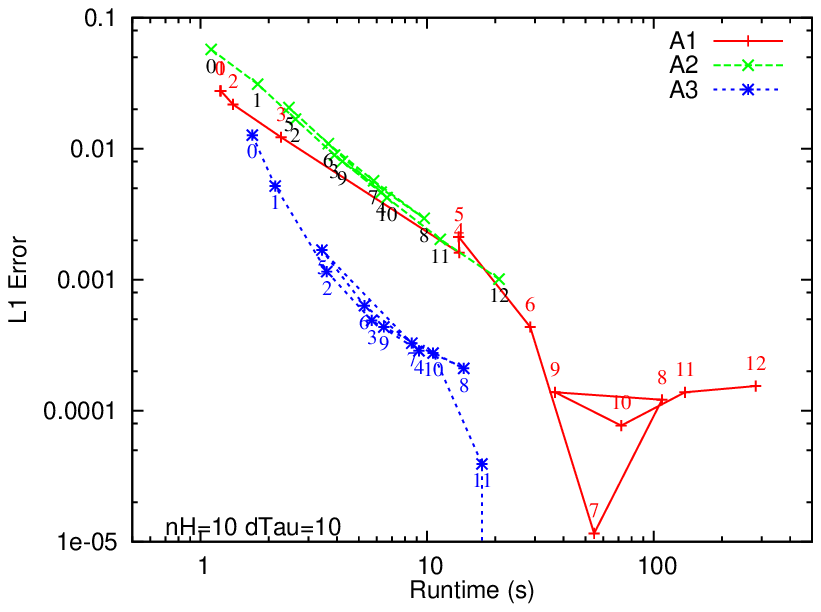}}
\resizebox{0.42\hsize}{!}{\includegraphics{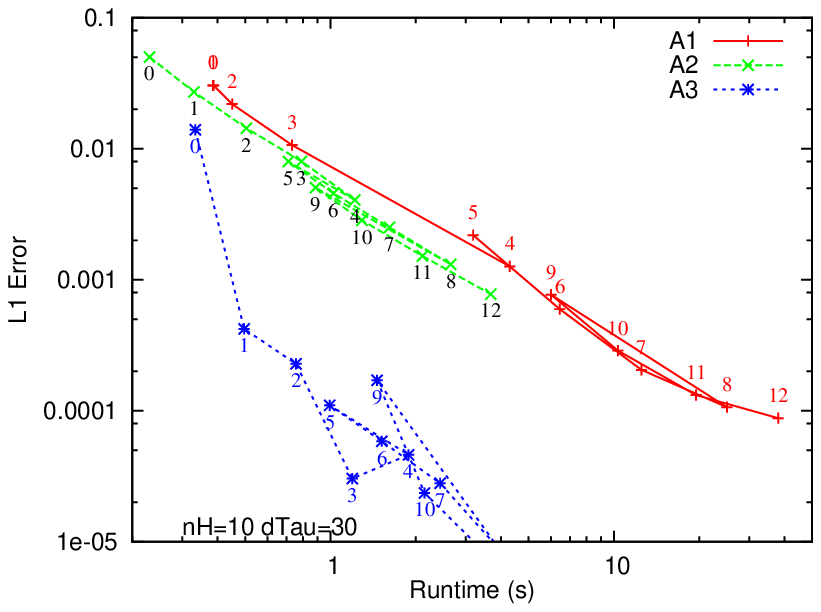}}
\caption{
  Accuracy as measured by the L1 error as a function of simulation runtime in seconds for the different algorithms and timestep criteria, for simulation outputs at $t=t_{\mathrm{rec}}$.
  The first three panels show results for the simulations with cell $\Delta\tau_0 \simeq1$ and ambient densities $n_{\mathrm{H}}=10\,\mathrm{cm}^{-3}$ (top left), $100\,\mathrm{cm}^{-3}$ (top right), and $1000\,\mathrm{cm}^{-3}$ (centre left).
  The next three show results for simulations with $n_{\mathrm{H}}=10\,\mathrm{cm}^{-3}$ and cell $\Delta\tau_0 \simeq3$ (centre right), $\Delta\tau_0 \simeq10$ (bottom left), and $\Delta\tau_0 \simeq30$ (bottom right).
  Each point represents a timestep criterion with A1 in red, A2 in green, and A3 in blue.
  The points are numbered according to the timestep criteria (A1 in red, A2 in black, A3 in blue), although not all numbers are readable due to overcrowding.
  }
\label{fig:MF_Efficiency}
\end{figure*}

All algorithms provide a numerically converged solution using the timestep criterion dt12, but this is unfeasibly restrictive for most problems.
The cases dt11 and dt12 bracket the criterion used by \citet{WhaNor06} and more recently by \citet{WisAbe11} but, as noted by \citet{WisAbe11}, a more efficient algorithm is desirable.
The data obtained here allow comparison of the numerical algorithms in terms of both speed and accuracy, which is shown in Fig.~\ref{fig:MF_Efficiency} by plotting the L1 error of the solution as a function of runtime (as a proxy for computational expense) at $t=t_{\mathrm{rec}}$.
The runtime of the simulations was calculated using a microsecond timer that starts once the code enters the main timestepping loop and stops when the code exits this loop.
The only data output during this time is for log-files (which is buffered) and, in addition, the calculations were run on a multi-core computer ensuring that at least one core was always idle.
The runtimes of some simulations are so short, however, that their accuracy needed verification.
To this end the code was instrumented with the \emph{Callgrind} tool of the \emph{Valgrind} profiling and debugging software suite\footnote{\url{http://valgrind.org}}.
This tool counts the instructions passed to the CPU and the results obtained were indistinguishable from Fig.~\ref{fig:MF_Efficiency} but with a rescaled $x$-axis, demonstrating that the runtimes in the figure are reliable.
The L1 error is here defined as the mean error per grid cell:
\begin{equation}
 \mathrm{Error} = \frac{1}{N_i}\sum_{i=0}^{N_i-1} |y_i-y_{i,\mathrm{ref}}| \,,
\end{equation}
where $y_{i,\mathrm{ref}}$ is the reference solution at cell $i$ of $N_i$ cells, taken as the A3-dt12 solution.
Note that this is not a relative error, so most of the contribution is from cells near the ionisation front where the differences in $y$ can be of the order of unity.

In all cases the A3 results lie below the A2 results, and the higher rate of convergence is also apparent, especially for calculations with high optical depth cells.
For calculations with $\Delta\tau_0 \simeq1$ A1 is more efficient than A3 using dt00-dt05, but these timestep criteria are not sufficiently accurate for $\Delta\tau_0 \gtrsim10$, and the timestep criteria dt06-dt12 are always more expensive with A1 than A3.
The convergence between the solutions for A1 and A3 levels off for the $\Delta\tau_0 \lesssim 3$ simulations (first four panels) at a relative difference of about $10^{-3}$.
The relative error tolerance in the microphysics integrator is set to $10^{-4}$ so it is not surprising that mean differences of $\lesssim10^{-3}$ are found.
For the same reason the fact that the solutions agree more closely (to $10^{-4}$) for $\Delta\tau_0 \simeq [10, 30]$ is probably not significant.

\subsection{Optimal timestep criteria}
The best timestep criterion for each algorithm is somewhat subjective, depending on what one considers to be an acceptable level of error compared to a fully converged solution.
Errors from discretisation in multi-dimensional simulations are generally $\sim1$\% so it seems reasonable to set the accuracy requirement at around this level.
In this case dt05 is sufficient for A1 in almost all situations, limiting the timestep by the relative change in internal energy and absolute change in $y$.
It also avoids the limitations of dt00-dt04, which limit $\Delta t$ by the recombination time even for an equilibrium situation where neither $y$ nor $E_{\mathrm{int}}$ is changing.
On the other hand, with dt05 an ionisation front takes a number of timesteps to cross a single cell, taking away the primary advantage A1 has over A3.

For A2, dt08, dt11 or dt12 give an acceptable level of accuracy in all situations, in agreement with the assessment of previous authors using similar criteria \citep{WhaNor06}.
Fig.~\ref{fig:MF_Efficiency} shows, however, that it is almost always the least efficient algorithm, and that A3 is a much better explicit integrator.
With A3, dt02 is already a good enough solution in all cases.
It is generally more accurate than dt05, although dt05 is superior for very optically thick grid cells.
Any of the criteria from dt02-dt12 are acceptable for A3, and if efficiency was not an issue dt03 or dt06 would be preferable.

\section{Parallel scaling}

\label{sec:scaling}
If we use the timestep criteria suggested in the previous section, then A3-dt02 is already the most efficient algorithm when run on a single core.
A3 should also scale efficiently to a larger number of cores because there are fewer calculations in the poorly scaling raytracing step.
In this case the motivation for comparing the scaling of the algorithms is not so much to demonstrate the advantage of one algorithm over the other, but rather to test the scaling of each algorithm individually.
It should be borne in mind that the overall algorithm efficiency is also strongly dependent on the microphysics integration algorithm, and a specially tailored scheme for A1 and A3 could probably be made more efficient than the generic backward-differencing integrator used here.
As an example, it is possible that the C$^2$-ray method of \citet{MelIliAlvEA06} is more efficient than the (similar) implementation used here as A1 and, if so, would have a lower normalisation on the following plots.
Raytracing here uses the short characteristics tracer that scales reasonably well on parallel architectures; the scaling may be different for other raytracing algorithms such as long characteristics which concentrate many rays near the source.
The following tests were performed on the JUROPA computer at the J\"ulich Supercomputing Centre in Germany.

The parallelisation strategy of the code is quite simple: the simulation domain is recursively divided into two $n$ times with the division being along the axis on which subdomains have the largest number of cells, resulting in $N=2^n$ subdomains of equal size that are as close to cubic as possible.
The radiation source is moved to the nearest cell vertex for raytracing purposes; its position can be chosen so that this vertex also lies on a subdomain vertex in which case quadrants/octants of the domain can be traced independently.
Each subdomain makes (for each source) a list of domains closer to the source from which it needs to receive boundary data column densities, and another list of the domains to which it has to send boundary data.
The boundary cells containing data to be sent and received are put into linked lists of pointers to cells for each boundary to be sent.
In the current implementation the subdomain face, edge, and corner data are sent separately, but they could (more simply) be sent together and this will be upgraded in the near future.
This upgrade will not change the scaling drastically, but should make the code a little more efficient.

The raytracing in parallel then consists of three functions.
The first function looks for boundary data to receive until it has received data for each boundary in the list.
Whenever data are received they are unpacked and copied into the relevant local boundary cells.
The second function calls the serial short characteristics raytracing routine on the local domain, and the third packages the outbound column density data and executes a non-blocking send for each subdomain it needs to send data to.

\begin{figure}
\centering
\resizebox{0.85\hsize}{!}{\includegraphics{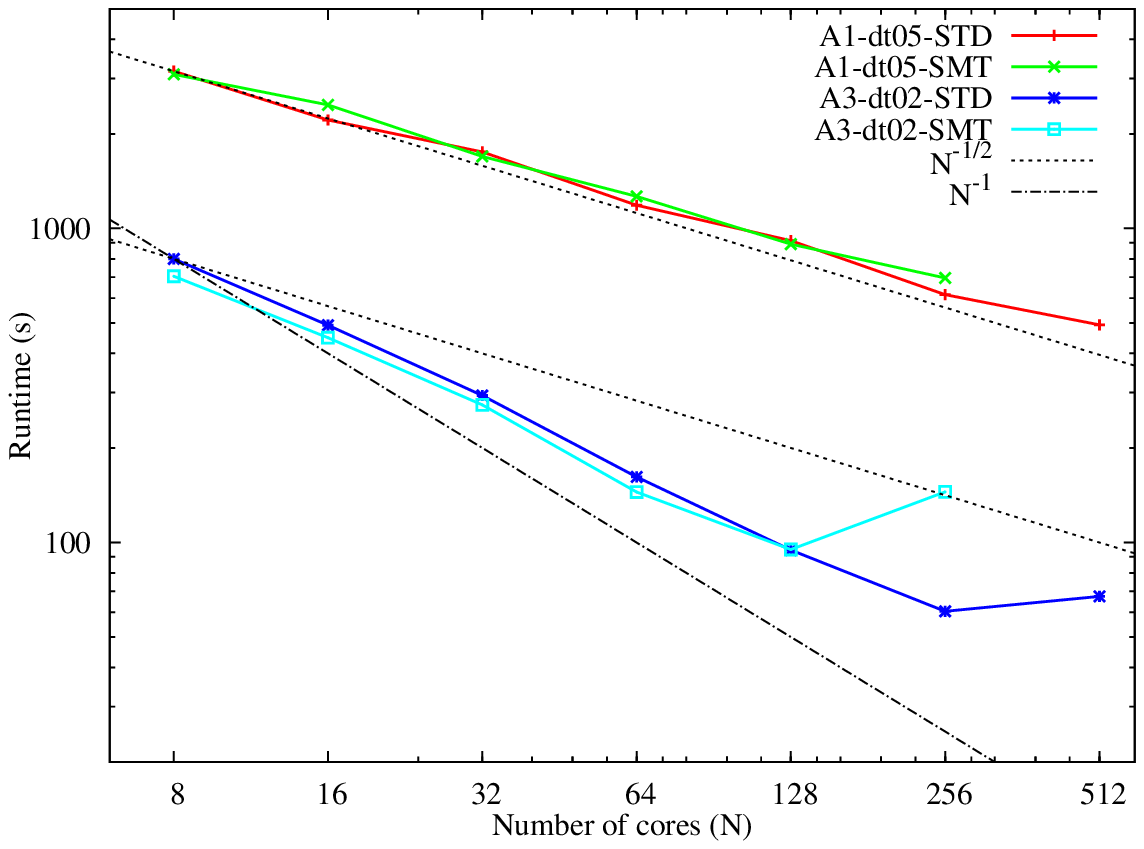}}
\resizebox{0.85\hsize}{!}{\includegraphics{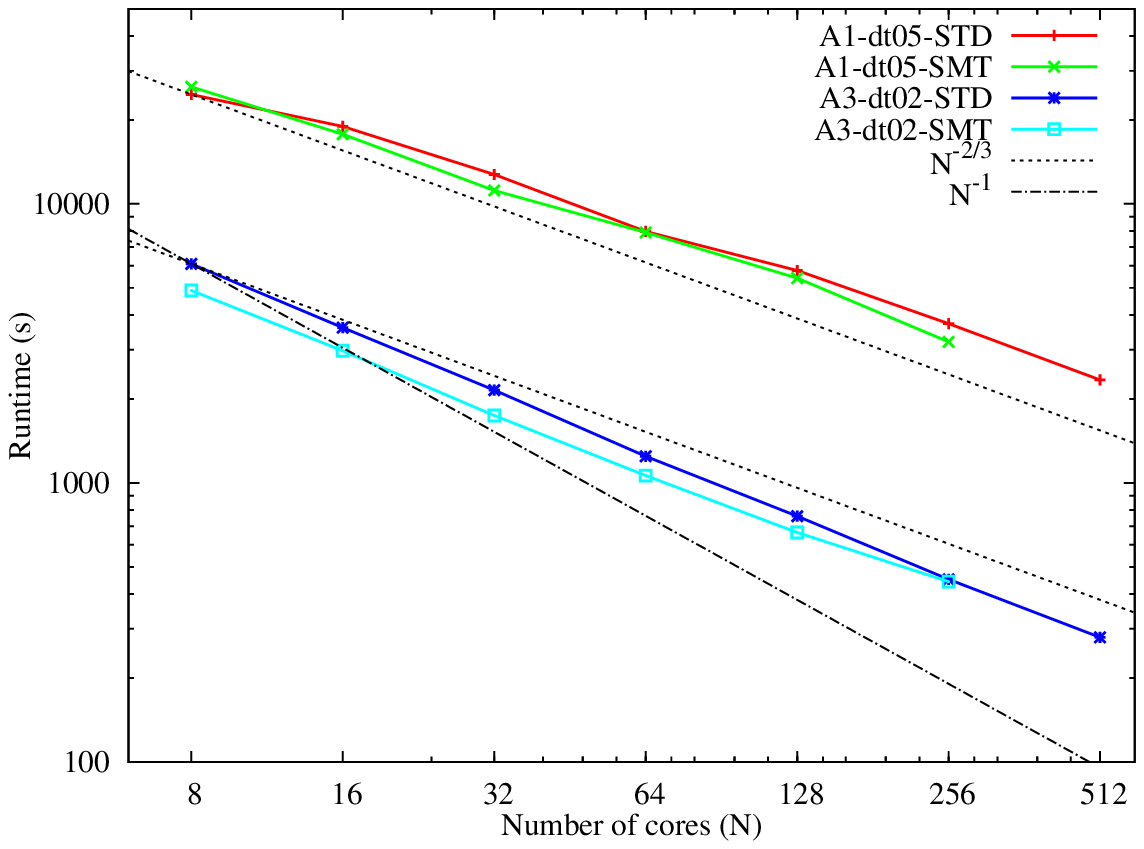}}
\caption{
  Scaling comparison between algorithms 1 and 3 on the distributed memory cluster JUROPA.  Above: Scaling of a 2D static photoionisation problem with $512^2$ grid cells as discussed in the text.  Below: Scaling of the same static problem, but now in 3D with $160^3$ grid cells.}
\label{fig:scaling}
\end{figure}

\subsection{Static 2D and 3D models}\label{sec:scaling_static}
Two simulations were performed of the expansion of an H~\textsc{ii} region into a uniform medium on a 2D axisymmetric grid and a 3D Cartesian grid.
The parameters are identical to that of the H~\textsc{ii} region in the previous section, using a multi-frequency ionising (and FUV heating) source; the 2D model has $512^2$ cells and the 3D model $160^3$, and only the positive quadrant/octant of the domain is simulated.
The model was run for 10 recombination times ($3.861\times10^{11}$ s), again with no dynamics.

The calculations were run first using one MPI process per physical core (denoted `STD'), and then repeated with two MPI processes per core using the \emph{simultaneous multi-threading} feature of JUROPA (denoted `SMT').
Both calculations were run using 8-512 MPI processes in STD mode, and 16-512 processes (on 8-256 cores) in SMT mode.
For calculations using 512 processes the subdomain for each process is $16\times32$ cells in 2D and $20^3$ cells in 3D.
These very small subdomains are unlikely to be used in a production calculation because the number of ghost/boundary cells is comparable to the number of real cells.

The strong scaling results are shown in Fig.~\ref{fig:scaling} in terms of the total wall-clock time to solution, where ideal scaling would have a slope of -1.
Raytracing using short characteristics must scale as $N^{-1/2}$ in 2D and $N^{-2/3}$ in 3D (where $N$ is the number of cores) in the limit of small subdomains and zero communication time, and this is clearly shown in the scaling of A1.
The reason for this scaling is that the rays must be traced causally outwards from the source, so parallelisation is always restricted in one of the spatial dimensions.
This means that for 2D calculations only a 1D curve of subdomains can be active at any time, and in 3D this is a 2D spherical shell of subdomains.
The scaling of the raytracing algorithm follows simply from this.

A1 scales as well as it can be expected to, but there seems to be little advantage in running 2D calculations on large numbers of cores using this algorithm.
The scaling stays roughly constant for A1 out to the maximum number of cores used, although there is an indication that the curve is flattening further at 256 and 512 cores.

The scaling of A3 should be better than that of A1 because the microphysics integrations are separate from the poorly scaling raytracing step, and so a larger percentage of the total computation can be performed fully in parallel.
For the 2D problem the scaling of A3 is indeed far superior to A1, and it is faster for all runs despite taking more timesteps (because of the different timestep criterion).
For the 2D problem with A3 there is no speedup in the raytracing when going from 64 to 128 cores with SMT and the overall calculation actually slows down for 256 cores (and for 512 cores in STD mode); this may be due to network latency becoming a limiting factor.
The total work is also being increased because the number of boundary cells is increasing to a large percentage of the total number of real cells, and it is more likely that this is the reason for the slowdown.
All processes write a small log-file, so another possibility is that the very short jobs with hundreds of cores are affected by disk I/O congestion.
The A1 calculations are running about $10\times$ slower, which means that issues such as network latency and disk I/O will not affect them to the same extent.
In 3D the difference in the scaling is much less significant, and if A1 could be made more efficient, it would be competitive with A3 even out to 512 cores.
Of course the prime motivation behind A1 was originally to enable R-type ionisation fronts to cross many cells accurately in a single timestep, and with dt05 this no longer happens, so it is unclear why one would continue using A1 unless it could be made sufficiently accurate with a less restrictive timestep.

Neither algorithm scales ideally (i.e.\ linear speedup with increasing core count); the reason for this is clear with A1, but for A3 it is also true when the raytracing is taking a negligible fraction of the runtime.
The reason for the less-than-ideal scaling in this case seems to be that the microphysics integrator does not have constant work per grid point, but instead the computation is concentrated near the ionisation front.
In ionised gas very little is changing, and in neutral gas this is also the case, so the microphysics integration is almost trivial.
Within the ionisation front, however, the equations are stiff and an accurate solution requires much more computation.
There is a boundary data exchange after the microphysics update, so effectively this step is limited by the slowest subdomain.
In principle this could be solved by active load-balancing, where the size of each subdomain is varied according to the computation time required.

\begin{figure}
\centering
\resizebox{0.85\hsize}{!}{\includegraphics{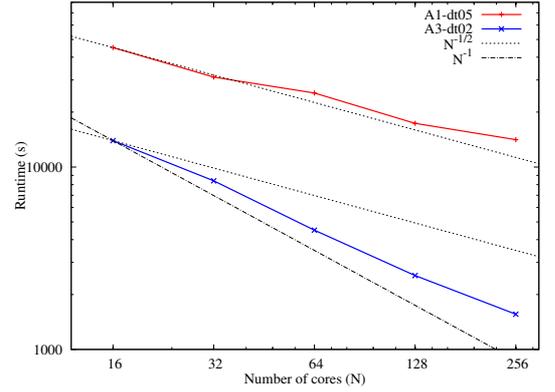}}
\caption{
  Scaling comparison between algorithms 1 and 3 on JUROPA for a 2D calculation with photoionisation and a $1000\,\mathrm{km}\,\mathrm{s}^{-1}$ stellar wind, as discussed in the text.}
\label{fig:scaling_dyn}
\end{figure}

\subsection{Models with a fast stellar wind}
To test the scaling in the opposite extreme where the dynamics strongly limits the timestep, a simulation was run including photoionisation and a stellar wind from a hot massive star.
The scaling results of this test are representative of a production calculation.
The star is moving with $4\,\mathrm{km}\,\mathrm{s}^{-1}$ through a constant density neutral medium with $n_{\mathrm{H}}=3000\,\mathrm{cm}^{-3}$, and is emits $3\times10^{48}$ ionising photons per second with a blackbody spectrum of $T=37\,500$K.
The stellar wind parameters are $\dot{M}=2\times10^{-7}\,\msun\,\mathrm{yr}^{-1}$ and $v_{\mathrm{w}}=1000\,\mathrm{km}\,\mathrm{s}^{-1}$.
An axisymmetric model was run with $512\times256$ grid cells and physical domain $z\in[-1.28,1.28]\times10^{18}$ cm and $R\in[0,1.28]\times10^{18}$ cm.
The wind was injected following \citet{vMarLanAchEA06} by imposing a freely expanding, adiabatically cooling, fully ionised wind within a radius of 15 cells of the origin (out to $0.75\times10^{17}$ cm).
The Str\"omgren radius was $R_S=6.75\times10^{17}$ cm, and in the initial conditions the ambient medium was ionised out to $1.92\times10^{17}$ cm, more than twice the radius of the wind boundary condition.
This substantially reduces the time during which the ionisation front is R-type, and consequently for most of the calculation the overall timestep was set according to the Courant condition on the $1000\,\mathrm{km}\,\mathrm{s}^{-1}$ wind region and not according to the microphysics timestep restrictions.
The simulation was run for 10 kyr, or $\simeq245t_{\mathrm{rec}}$.
The scaling results in Fig.~\ref{fig:scaling_dyn} (this time only for STD mode with one MPI process per physical core) show similar results to the static 2D case for A1.
The scaling of A3 is somewhat better than for the static case, probably because each timestep now requires more computation and hence fewer timesteps are completed per second for a given number of cores.
Indeed, the speedup is basically linear with A3 from 32 to 128 cores and, in terms of total core-hours required, the 256 core run still has an efficiency of 56\% compared to the 16 core run.

\section{Discussion and conclusions} \label{sec:conc}
\subsection{Explicit algorithms}
Variants of three commonly used photoionisation tracking algorithms have been implemented and tested for their accuracy, efficiency and scaling properties.
The first-order accurate explicit algorithm, A2, was shown to be the least accurate and efficient when R-type ionisation fronts are present, with one of A2-dt08, dt11, or dt12 required for reasonably converged solution.
Algorithm 3 (A3) is an extension of A2 to second-order time accuracy; its advantages over A2 in terms of photon conservation and efficiency for different timestepping criteria are presented here for the first time.
The results of this comparison demonstrate that A2 should always be rejected in favour of A3 (or a similar higher order integration) when an explicit algorithm is to be used.
A sufficient timestep criterion for photon conservation and ionisation front tracking is A3-dt02, where the microphysics timestep limit is $\Delta t=0.25/\dot{y}$.
This could possibly be relaxed to $\Delta t=0.5/\dot{y}$ if errors of $\sim5$ per cent are considered acceptable, but the errors are $>10$ per cent (and fairly unpredictable) if $\Delta t=1.0/\dot{y}$ is used.
The performance of this timestep criterion in dynamical multi-dimensional simulations will be tested in future work.
Both A2 and A3 improve in accuracy with higher cell optical depths, but the gain is more noticeable with A3.
These conclusions comparing A2 to A3 hold for both monochromatic and multi-frequency ionising radiation.

In comparison with this work, the algorithm of \citet{WhaNor06} is closest to A2 in that it uses instantaneous column densities and is first-order accurate in time (for photon conservation).
These authors split the time-integration into two steps: a full timestep where the dynamics and microphysics are updated, limited by the Courant condition and by the condition that the internal energy in any cell can change by at most 10\%; and also a substep in which the raytracing is performed and the chemical network updated, with the sub-timestep set by the requirement that the electron density change by at most 10\%.
Each full timestep therefore consists of one or more substeps.
Similar timestep limits are also used by \citet{KruStoGar07} and \citet{WisAbe11}, with both algorithms closely related to A2 based on the published description.
This timestep criterion is quite close to dt11, where the internal energy is allowed to change by at most 12.5\% and the neutral fraction by at most 12.5\% (which is approximately the inverse of the electron fraction).
Using A2, it is shown here that dt11 and dt12 (as well as dt08) give basically converged results in all situations, in agreement with \citet{WhaNor06}.
In addition, it has been shown here that these criteria are sufficient for multi-frequency radiation as well as the monochromatic radiation used in these previous works.

The clear advantages of A3 over A2 in terms of accuracy and efficiency suggest, however, that other codes could make large efficiency gains by switching to an algorithm similar to A3-dt02 for the raytracing/chemistry step (or substep).
Although this involves two raytracings and chemistry integrations per (sub)step instead of one, the much longer timestep allowed means that less total computation is required, and furthermore that fewer raytracings are required for a given computation.
This is important because raytracing is the major bottleneck for the parallelised AMR implementation of \citet{WisAbe11}.
The improvement in A3 compared to A2 is independent of spatial resolution, requiring only a higher order, finite volume formulation, so it is expected that the same gains in efficiency and accuracy presented here could just as easily be gained by AMR codes, although the parallel scaling is admittedly much more complicated.
A3 can also be applied to photoionisation substeps by simply omitting the dynamics updates from the sequence of steps shown in Fig.~\ref{fig:Algorithm3}.
While the mass density field is not time-centred in a substep, the ion fractions and hence the column densities are, and if substepping is employed it means the density field is evolving on a much longer timescale anyway.

Even with the second-order accurate A3, errors greater than 10 per cent are obtained using dt00 with $K_4=1$; this likely explains why the \citet{RijPleDubEA06} \textsc{flash-hc} algorithm had difficulty tracking R-type ionisation fronts accurately in the code comparison project of \citet{IliCiaAlvEA06}.
As an explicit algorithm, it is not possible to accurately track R-type ionisation fronts crossing more than one optically thick cell per raytracing.
On the other hand, D-type ionisation fronts are subsonic by definition (with respect to at least one of the gas components), and so the Courant condition automatically imposes $K_4<1$, leading to accurate ionisation front propagation.

\subsection{Implicit algorithm}
Implicit methods have, in principle, a clear efficiency advantage over explicit methods for R-type ionisation fronts, at least for monochromatic radiation, although the timestep does need to be restricted to a fraction of $t_{\mathrm{rec}}$.
For A1, which is similar but not identical to the C$^2$-ray method of \citet{MelIliAlvEA06}, ionisation fronts are tracked with negligible error for tests with monochromatic radiation and no recombinations, shown in section~\ref{sec:acc}.
When recombinations are switched on the accuracy is somewhat lower, but the expansion of an H~\textsc{ii} region is tracked by a single timestep to $t=0.1t_{\mathrm{rec}}$ with $<1.5$\% error (with criterion dt02).
For this reason A1-dt02 is a vastly more efficient algorithm than A2 or A3 for tracking R-type ionisation fronts with monochromatic radiation.

The situation is not so simple with multi-frequency radiation.
For a given emitted radiation spectrum, the transmission of photons through a cell depends not only on the optical depth of the cell, but also on the optical depth from the source to the cell (because this changes the incident radiation spectrum).
In addition, when the optical depth within a cell changes by a value $\gg1$ over a timestep, the transmitted spectrum also changes significantly, and it is no longer clear that a time-averaged optical depth (or equivalently attenuation fraction at a specific frequency) will give good photon conservation.
This is borne out in the results from section~\ref{sec:multifrequency} where A1 performs well when cell optical depths are $\Delta\tau_0 \lesssim3$, but the accuracy decreases as $\Delta\tau_0$ increases for dt00-dt04.
To achieve good accuracy for all densities and values of $\Delta\tau_0$ one of dt05-dt12 must be used, removing entirely the efficiency advantage of A1 over A3.
Here it is suggested that dt05 represents the best balance of accuracy and efficiency, although it remains significantly less efficient than A3-dt02.

No attempt has been made to modify A1 to produce better results with multi-frequency radiation.
It is possible that choosing more carefully the frequency at which the time-averaged attenuation fraction is calculated would give a more accurate result, and it is also possible that modifying the time-averaging strategy would improve photon conservation.
It is also possible that the C$^2$-ray method (which does have a different time-averaging strategy) is already more accurate than A1 with multi-frequency radiation.
Testing these hypotheses is, however, beyond the scope of this work.

\subsection{Parallel scaling}
In the limit of many processors with small simulation domains the time for the raytracing step scales with $N^{-1/2}$ in 2D calculations and $N^{-2/3}$ in 3D (where $N$ is the number of cores) when using the method of short characteristics to trace rays from point sources in Cartesian geometry.
This is because the rays must be traced outwards from the source in sequence, so only a 2D surface (or 1D curve) of subdomains can be active at any time in a 3D (or 2D) raytracing.
The scaling of A1 closely follows these power laws, both for static and dynamical simulations.
A3 scales significantly better than A1 in 2D calculations and somewhat better in 3D; this is a consequence of having a smaller fraction of the total computation in the raytracing step.
The scaling of A3 is less than ideal even when raytracing is not a limiting factor, and this is likely due to imbalances in the work required for the microphysics integration (where most computation is required near the ionisation front).
Despite this, the efficiency of A3 in 2D and 3D remains above 50\% on up to $N>100$ cores for the test calculations run here (up to 256 cores for the test including dynamics).

The scaling of A2 should be the same as A3, but it is less efficient and less accurate, so it was not tested here.
The result that A3 is here always more efficient than A1 for all $N$ is certainly implementation-dependent, and if a version of A1 could be devised that allows (for example) timestep criterion dt02 to be used, then A1 would suddenly become much more efficient by virtue of requiring many fewer timesteps than A3.
For general problems, however, A3 has a higher order of accuracy and better scalability than A1 and so may be more suitable for parallel simulations where photoionisation has a strong effect on gas dynamics.
In addition, the simplest possible non-equilibrium chemistry model has been used here; the scaling advantage of A3 should be more important when more computation is required in the chemistry/thermal physics source term integration by e.g.\ the inclusion of He and H$_2$.

\begin{acknowledgements}
The author is grateful to Sam Falle for pointing out the benefits of the integration scheme used in A3; to Thomas Peters for useful discussions regarding the \textsc{flash-hc} algorithm and his extensions to it; and to the referee Garrelt Mellema for insightful suggestions which substantially improved this paper.
This work was part funded by an Argelander Fellowship and by a fellowship from the Alexander von Humboldt Foundation.
The author acknowledges the John von Neumann Institute for Computing for a grant of computing time on the JUROPA supercomputer at J\"ulich Supercomputing Centre.
\end{acknowledgements}

\vspace{-0.5cm}
\bibliographystyle{aa}
\bibliography{jmackey}
\end{document}